\documentclass[fleqn]{aa}  
\usepackage{supertabular,booktabs}  % For the big table
\usepackage{graphicx}
\usepackage{xcolor}
\usepackage{hyperref}
\usepackage{txfonts}
\usepackage{textgreek}
\usepackage{calligra}
\usepackage{calrsfs}
\usepackage[T1]{fontenc}
\usepackage[mathcal]{eucal}
\usepackage{longtable}
\usepackage{multicol}
\usepackage{subcaption}

\usepackage[switch]{lineno}
\sfcode`A=1000 \sfcode`B=1000 \sfcode`C=1000 \sfcode`D=1000
\sfcode`E=1000 \sfcode`F=1000 \sfcode`G=1000 \sfcode`H=1000
\sfcode`I=1000 \sfcode`J=1000 \sfcode`K=1000 \sfcode`L=1000
\sfcode`M=1000 \sfcode`N=1000 \sfcode`O=1000 \sfcode`P=1000
\sfcode`Q=1000 \sfcode`R=1000 \sfcode`S=1000 \sfcode`T=1000
\sfcode`U=1000 \sfcode`V=1000 \sfcode`W=1000 \sfcode`X=1000
\sfcode`Y=1000 \sfcode`Z=1000

\begin{document} 
% \linenumbers

   \title{Gaussian Process Inference of Stochastic Magneto-Active Dynamics and Viscosity in Swift J1727.8-1613}

   % \titlerunning{Gaussian Process Inference of Stochastic Magneto-Active Dynamics and Viscosity in Swift J1727.8-1613}
   \author{Lijuan Dong \inst{1}
       \and Dahai Yan \inst{1} \fnmsep\thanks{yandahai@ynu.edu.cn}
       \and Zihan Yang \inst{2}  
       \and Haiyun Zhang \inst{3}
       \and Lin Xie \inst{1}
       \and Qingcui Bu \inst{4}
       \and Lian Tao \inst{2}
    }      
   \institute{Department of Astronomy, School of Physics and Astronomy, Yunnan University, Kunming, Yunnan, 650091, People’s Republic of China 
        \and Key Laboratory for Particle Astrophysics, Institute of High Energy Physics, Chinese Academy of Sciences, 19B Yuquan Road, Beijing 100049, China 
        \and Department of Physics, School of Physics and Astronomy, Yunnan University, Kunming, Yunnan, 650091, People’s Republic of China
        \and Institute of Astrophysics, Central China Normal University, Wuhan 430079, People’s Republic of China 
          }
 \abstract{
  Linking X-ray variability to the underlying magnetohydrodynamic (MHD) dynamics of black hole X-ray binaries remains challenging. We systematically investigate the stochastic and oscillatory variability of the black hole X-ray binary candidate Swift J1727.8$-$1613 during its 2023 outburst using Gaussian process (GP) regression applied to Insight-HXMT multi-band light curves. The variability is modeled with a physically motivated composite kernel comprising one stochastically driven damped simple harmonic oscillator (SHO) and two damped random walk (DRW) components. The SHO term robustly recovers quasi-periodic oscillations (QPOs) with frequencies $\nu_0 \sim 0.07$--$5$ Hz, consistent with the fundamental Alfv\'en mode of a contracting magnetically confined disk--coronal cavity. The quality factor rises from $Q \sim 3$ to $Q \sim 10$, suggesting increasing coherence of the magnetic cavity. We also find an anti-correlation between QPO frequency and the short DRW damping timescale, supporting our proposed stochastic magneto-active dynamics scenario. We associate the short and long DRW timescales with the local turbulent turnover and thermal adjustment timescales, respectively. Under this proposed interpretation, our results are consistent with an effective viscosity parameter $\alpha \sim 0.1$. Strikingly, near the jet-ejection epoch around MJD 60206, the short DRW timescale is unresolved, while the long DRW timescale approaches the 0.1 s sampling limit, suggesting a rapid reorganization of the disk internal energy balance immediately before jet launching. Our results establish GP inference as a powerful route to connecting X-ray timing observables with the dynamical state of black hole accretion flows.

   }   
   \keywords{Methods: data analysis – Methods: statistical – X-rays: binaries
               }

   \maketitle
%
%-------------------------------------------------------------------
\section{Introduction}
\label{s:1}
%%%%%%%%%%%%%%%%%%%%%%%%

%Intro Swift J1727.8−1613

Swift J1727.8-1613 was first detected by Swift/BAT and MAXI/GSC on 24 August 2023, and was subsequently identified as a candidate black hole X-ray binary (BHXRB) based on its exceptionally high X-ray flux and the presence of canonical spectral state transitions \citep{2023GCN.33465....1B,2024MNRAS.529.4624Y,2024ApJ...966L..35S}. 
The black hole binary nature of the system was dynamically confirmed through optical spectroscopy, which measured an orbital period of $10.804 \pm 0.001$ hours \citep{2025A&A...693A.129M}. 
Based on the measured radial velocity semi-amplitude of the donor star ($K_2 = 390 \pm 4$ km s$^{-1}$), the mass function was derived to be $f(M) = 2.77 \pm 0.09 M_{\odot}$. This implies a lower limit on the compact object mass of $M_1 > 3.12 \pm 0.10 M_{\odot}$, thereby excluding a neutron star primary \citep{2025A&A...693A.129M}. 
The companion star is classified as a K4V main-sequence star; however, its spectral contribution appears to be significantly obscured by emission from the accretion disk \citep{2025A&A...693A.129M}. 
Regarding the distance to the system, dynamical constraints suggest $d\approx3.4 \pm 0.3$ kpc \citep{2025A&A...693A.129M}, whereas independent estimates based on neutral-hydrogen absorption and ultraviolet spectroscopy favor a larger distance of $d\approx5.5^{+1.4}_{-1.1}$ kpc \citep{2025ApJ...994..243B}.

%Intro disk
The inner disk radius in the hard state remains debated. In the truncated-disk scenario described by \cite{1997ApJ...489..865E}, the optically thick disk lies outside the ISCO in the LHS and moves inward as the source softens. However, \cite{2013MNRAS.431.3510S} showed that spectral hardening can mimic changes in the inferred inner radius, while \cite{2015ApJ...813...84G} found that the disk may extend close to the ISCO even in the hard state. For Swift J1727.8$-$1613, \cite{2026ApJ..1003...73K} likewise emphasized the effects of model systematics on disk-radius estimates. \cite{2026ApJ..1000..137C} inferred a truncated disk in the HIMS, with $R_{\rm in}\gtrsim33 r_{\rm g}$, followed by an inward shift in the SIMS to $R_{\rm in}\lesssim2.6 R_{\rm ISCO}$. This evolution was accompanied by increases in the QPO frequency from $\sim1.3$ to $\sim6.6$ Hz and in the disk temperature from approximately $0.4$ to $0.9$ keV.
\cite{2026ApJ..1000..137C} also reported an increase in the fitted disk electron density from $n_{\mathrm{e}} \sim 10^{16}~\mathrm{cm}^{-3}$ to $10^{17}~\mathrm{cm}^{-3}$ from the HIMS to the SIMS, although the SIMS value was not independently constrained. These densities remain well below those expected for BHXRB disks, so their absolute values and evolution should be interpreted cautiously due to reflection-model limitations \citep{2026NewAR.10201746Z}.

%Intro corona
During the LHS, the corona emission is dominated by Comptonized hard X-ray emission, with thermal Comptonization as the primary radiation mechanism. The disk provides soft seed photons for Comptonization in the corona, which irradiates the disk and produces detectable reflection features. Both components evolve synchronously with changes in the accretion rate \citep{2023ApJ...958L..16V,2024ApJ...968...76I, 2024ATel16541....1P, 2025ApJ...993...40X}. In the HIMS, modeling requires two physically distinct Comptonizing regions. The hard X ray tail above 100 keV indicates a hybrid electron distribution, with the electron temperature of the hard region at approximately 13-18~keV \citep{2026ApJ..1000..137C}. In contrast, during the SIMS, the corona simplifies to a single Comptonizing region, characterized by a steeper photon index, a reduced covering fraction, and a weakened overall radiation contribution \citep{2026ApJ..1000..137C}. The corona is inferred to be extended, rather than spherical, and is preferentially oriented along the accretion-disk plane, i.e., perpendicular to the jet axis. This geometry is supported by the agreement between the X-ray polarization angle and the sub-mm polarization angle, which traces the jet direction \citep{2023ApJ...958L..16V, 2024A&A...686L..12P}. 

%Intro jet
Radio observations reveal that Swift~J1727.8$-$1613 hosts an unusually extended, two-sided, continuous jet during the HS and HIMS. The jet is oriented along the north–south axis and constitutes the most spatially resolved continuous jet observed in an X-ray binary \citep{2024ApJ...971L...9W, 2025ApJ...983...23C}. A key event took place at MJD~60206 (19 September 2023), coincident with the peak of a soft X-ray flare. At this time, the source briefly transitioned from the HIMS to the SIMS \citep{2025arXiv251010353J}. This transition was accompanied by three observational signatures. First, three discrete radio jet knots were ejected simultaneously \citep{2025ApJ...984L..53W}. Second, the broadband X-ray noise in the 2-10~keV band dropped sharply. Third, a Type-B QPO appeared, characterized by a quality factor $Q \geq 6$ and a U-shaped phase-lag spectrum, while the previously persistent Type-C QPO temporarily disappeared \citep{2025arXiv251010353J}. The anomalous radio-quiet radio-X-ray correlation and its break at the spectral softening point further show that accretion flow evolution, rather than jet physics alone, governs the state transition \citep{2025ApJ...988..109H, 2025MNRAS.542.1803H}. Dynamical modeling links the jet ejection at MJD~60206 directly to changes in the X-ray timing properties, including a rapid increase in QPO frequency \citep{2025ApJ...984L..53W, 2025arXiv251010353J}. 
The radio-loud behavior during this interval is consistent with the often proposed association between Type-B QPOs and discrete jet ejections in black hole X-ray binaries \citep{2009MNRAS.396.1370F, 2012MNRAS.421..468M}. However, this association is not universally observed and remains debated, as illustrated by the counterexample reported by \cite{2024MNRAS.533.4188C}.

The Lense-Thirring (LT) precession model attributes QPOs to global precession of the inner accretion flow caused by frame dragging around a rotating black hole. Although this model successfully predicts the relation between the QPO frequency and the disk truncation radius, it does not naturally explain the persistence of a high quality factor \citep{2009MNRAS.397L.101I, 2025MNRAS.543.1748M}. The accretion-ejection instability model invokes spiral instabilities in magnetized disks, which generate Rossby vortices that can modulate the emission and produce QPOs \citep{2002A&A...387..487R}. However, phase-resolved polarimetric studies of low-frequency QPOs showed no significant modulation of either the polarization degree or angle with QPO phase. This result is inconsistent with predictions from the LT precession model \citep{2024ApJ...961L..42Z}. The main obstacle to distinguishing between these models is the lack of a formal framework that directly links light-curve statistics with the underlying magneto-hydrodynamic (MHD) dynamics. 

\citet{2022ApJ...936...17H} established an important precedent for GP-based QPO searches in astrophysical transients. They modeled QPOs as stochastic processes superposed on deterministic transient profiles and used Bayesian model comparison to distinguish QPO signals from red-noise alternatives. More recently, \citet{2025A&A...703A.134W} applied a Gaussian Process method to the X-ray light curves of MAXI J1348-630. They used a composite kernel of a stochastically driven damped simple harmonic oscillator (SHO), a damped random walk (DRW), and an additional white noise (AWN). Their results captured the QPO signal with a period of 0.2 s, while $Q$ remained stable at $\sim10$. Their results revealed a hierarchy of timescales in the accretion flow. The AWN component is interpreted as a rapid DRW, which may trace high-frequency fluctuations that occur in both the inner region of the accretion disk and the corona. 
The rapid AWN-like component may be related to inward-propagating accretion-rate fluctuations, as discussed by \cite{1997MNRAS.292..679L, 2006MNRAS.367..801A}.

In contrast to the timing-analysis methods that operate in the frequency domain, such as power spectral density (PSD) modeling, GP inference offers a robust framework for analyzing aperiodic and quasi-periodic variability directly in the time domain \citep{2021ApJ...919...58Z,2022ApJ...930..157Z}. The primary advantage of the GP approach lies in its ability to decompose complex light curves into distinct, physically motivated stochastic processes through the selection of specific kernel functions \citep{2025MNRAS.537.2380Z, 2025A&A...703A.134W}. Furthermore, by utilizing specific GP kernels that correspond to linear stochastic differential equations (SDEs), this framework allows for the direct extraction of relaxation timescales and quality factors. These parameters provide a more intuitive mapping to the underlying magneto-hydrodynamics of the accretion flow compared to the integrated power or broad-band noise components typically measured in the Fourier domain. This makes such physically motivated GP modeling an ideal tool for probing the rapid dynamical transitions and coupling between different physical scales during black hole X-ray binary outbursts.

In this work, we present a systematic timing analysis of the 2023 outburst of the BHXRB Swift~J1727.8-1613 using multi-band observations obtained with Insight-HXMT. We characterize the X-ray variability with a composite kernel consisting of one SHO component and two DRW components. Our analysis reveals that the QPO frequency $\nu_0$ increases from $\sim 0.07$ to $\sim 5\,\mathrm{Hz}$, which is interpreted as the contraction of a magnetically confined resonant cavity. 
Simultaneously, $Q$ evolves from $\sim 3$ to $\sim 10$, indicating a transition toward a more coherent and rigid magnetic structure. 
Under our proposed identification of the short and long DRW timescales with the dynamical and thermal timescales of the accretion disk, respectively, the ratio $\tau_{\rm short}/\tau_{\rm long}$ is broadly consistent with an effective viscosity parameter of order $\alpha \sim 0.1$. 

This paper is organized as follows. Section~\ref{s:2} describes the data reduction and analytical methods. Section~\ref{s:3} presents the results of the GP fitting. In Section~\ref{s:4}, we propose a stochastic magneto-active dynamics scenario. Finally, the conclusions are presented in Section~\ref{s:5}.

\section{Data reduction and analytical methods}  
\label{s:2}   

\subsection{Data reduction}  
%Intro about data reduction
%Fig_1-1.pdf
\begin{figure} 
  \includegraphics[width=\linewidth]{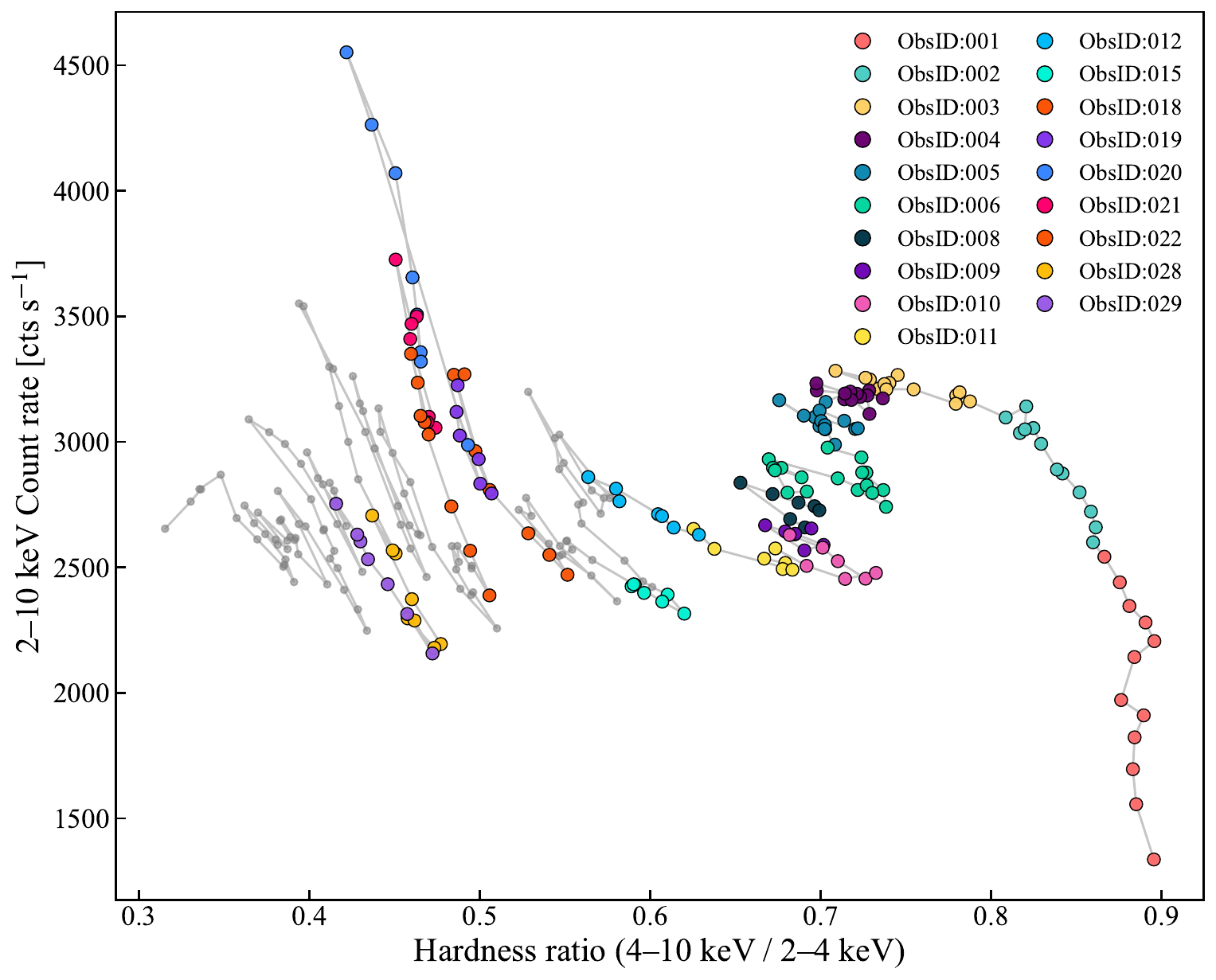}   
      \caption{The HID of Swift J1727.8-1613 during the 2023 outburst. The Y-axis represents the LE count rate in the 2--10~keV energy range, while the X-axis shows the hardness ratio, defined as the ratio of the count rates in the 4--10~keV and 2--4~keV bands. Colored circles highlight specific observation IDs (ObsIDs) selected for detailed timing analysis.}
         \label{Fig1}
\end{figure}

Launched on 2017 June 15, Insight-HXMT is China’s first X-ray astronomical satellite and the country’s first dedicated X-ray observatory. It aims to observe X-ray sources across a broad energy band spanning 1–250 keV \citep{2020SCPMA..6349502Z}. The Insight-HXMT consists of three payloads: Low Energy X-ray Telescope (LE, 1-15 keV; \cite{2020SCPMA..6349505C}), Medium Energy X-ray Telescope (ME, 5-30 keV; \cite{2020SCPMA..6349504C}), and High Energy X-ray Telescope (HE, 20-250 keV; \cite{2020SCPMA..6349503L}). We use the Insight-HXMT data analysis software Insight-HXMT (version 2.06) for HE and ME to process the data, while the LE data are filtered according to the criteria recommended in the Insight-HXMT Data Reduction Guide v2.07. Here we analyze Insight-HXMT data of Swift J1727.8-1613 from observation IDs: P0614338001-12, P0614338015, P0614338018-22, P0614338028-29, which were selected for model fitting analysis. We generate good time intervals (GTIs) using the following standard criteria: a pointing offset angle of less than $0.04^\circ$, an Earth elevation angle larger than $10^\circ$, a geomagnetic cutoff rigidity greater than 8 GV, and at least 300 s before and after passage through the South Atlantic Anomaly. To avoid contamination from the bright Earth and nearby sources, only data from the small FoV detectors were used.

Fig.~\ref{Fig1} shows the evolution of the hardness–intensity diagram (HID) across different observations. The hardness ratio (4–10 keV / 2–4 keV) decreases from 0.9 to 0.3, while the 2–10 keV count rate varies between 1500 and 4500 cts s$^{-1}$. The source follows the initial segment of the characteristic counter-clockwise track of black hole transients, starting from the rightmost hard state, progressing through the intermediate states, and moving toward the upper left at the peak of the outburst. 
The distribution in the HID is complex and does not delineate a complete q-shaped track. 
Nevertheless, the overall progression from the hard state toward softer intermediate states is broadly consistent with that previously reported for Swift J1727.8-1613 by \cite{2024MNRAS.529.4624Y} and  \cite{2025ApJ...986....3L}. The observations analyzed in this paper are marked with points of different colors. 

Fig.~\ref{Fig2} presents the temporal evolution of X-ray count rates in three energy bands.
The ME band reaches its peak at MJD 60184.12 with a count rate of
$\sim1300~\mathrm{cts~s^{-1}}$, while the HE band peaks earlier at $\sim3500~\mathrm{cts~s^{-1}}$. 
Both the ME and HE bands show a steady downward trend, with only minor fluctuations superposed on the decline.
The count rates drop to $\sim200~\mathrm{cts~s^{-1}}$ and $\sim500~\mathrm{cts~s^{-1}}$, respectively.
In contrast, the LE band shows a different behavior. It rises from MJD 60182 to 60185, reaching a plateau of $\sim3200~\mathrm{cts~s^{-1}}$ that lasts until MJD 60190. After a slight decrease, the source enters a flaring state around $\mathrm{MJD} \gtrsim 60197$. During this phase, the LE band displays intense variability with multiple rapid flares. A major flare occurs near MJD 60206, where the count rate sharply peaks at $\sim4500~\mathrm{cts~s^{-1}}$. This peak coincides with the successive ejections of jet knots 3, 2, and 1 between MJD 60206.22 and 60206.41 \citep{2025ApJ...984L..53W}.

\subsection{Gaussian process method}
%Gaussian process
GP regression is a non-parametric Bayesian method which has been used in time-domain astronomy \citep{2023ARA&A..61..329A}. 
%Its primary advantage is the ability to decompose light curves directly in the time domain. 
%It provides a flexible framework for modeling light curves, enabling robust extraction of variability patterns and characteristic timescales with clear physical interpretability \citep{2022ApJ...930..157Z,2025MNRAS.537.2380Z, 2025A&A...703A.134W}. 
In probability theory and statistics, GP is a stochastic process defined as a distribution over functions on a continuous domain, such as time or space. Based on Gaussian distributions, a GP can be regarded as an infinite-dimensional generalization of the multivariate normal distribution. For any finite set of input points $\{x_i\}_{i=1}^N$, the GP yields the function values $\boldsymbol{y} = (y_1, \dots, y_N)^T$. It follows a multivariate Gaussian distribution, $p(\mathbf{y})=\mathcal{N}(\mathbf{m},\mathbf{K})$, where $\mathbf{m}$ is the mean vector and $\mathbf{K}$ is the covariance matrix.

GP methods have successfully characterized QPOs and stochastic variability in active galactic nuclei \citep{2021ApJ...919...58Z, 2022ApJ...930..157Z, 2023ApJ...944..103Z, 2025ApJ...988..206Z, 2025MNRAS.540.3790Z}. The \texttt{celerite} model employs kernels composed of complex exponential mixtures, yielding a semiseparable covariance matrix. This property allows efficient factorization with $\mathcal{O}(N J^{2})$ complexity, where $N$ is the number of data points and $J$ is the number of kernel components, thereby alleviating the computational bottleneck in large astronomical time-series analyses \citep{2017AJ....154..220F, celerite2}. The covariance function is expressed as \citep{1995PhRvL..74.1060R}:
\begin{equation}
    \begin{split}
    k_{\boldsymbol{\alpha}}(\tau_{nm}) &= \sigma_{n}^{2} \delta_{nm} + \sum_{j=1}^{J} \left[
        \frac{1}{2}(a_{j} + i b_{j}) \exp \left( -(c_{j} + i d_{j}) \tau_{nm} \right)  + \right. \\
        &\quad \left. \frac{1}{2}(a_{j} - i b_{j}) \exp \left( -(c_{j} - i d_{j}) \tau_{nm} \right) \right] \\
    &\equiv \sigma_{n}^{2} \delta_{nm} + \sum_{j=1}^{J} \left[ 
        a_{j} \exp \left( -c_{j} \tau_{nm} \right) \cos \left( d_{j} \tau_{nm} \right) + \right. \\
        &\quad \left. b_{j} \exp \left( -c_{j} \tau_{nm} \right) \sin \left( d_{j} \tau_{nm} \right) \right],
    \end{split}
\label{covariance function}  
\end{equation}
where $\boldsymbol{\alpha}$ denotes the vector of hyperparameters, $\tau_{nm} = |t_n - t_m|$ is the time difference, $\{\sigma_n^2\}_{n=1}^N$ are the measurement uncertainties, and $\delta_{nm}$ is the Kronecker delta function. The parameters $a_j$ and $b_j$ are amplitude terms, $c_j$ controls the exponential decay of the oscillation, and $d_j$ is the oscillation frequency. The Fourier transform of Eq.~\ref{covariance function} is the PSD of the process, and it is obtained as
\begin{equation}S(\omega)=\sum_{j=1}^J\sqrt{\frac{2}{\pi}}\frac{(a_jc_j+b_jd_j)\left(c_j^2+d_j^2\right)+(a_jc_j-b_jd_j)\omega^2}{\omega^4+2\left(c_j^2-d_j^2\right)\omega^2+\left(c_j^2+d_j^2\right)^2}.
\label{q:3}
\end{equation}

When $b_j = 0$ and $d_j = 0$, the covariance function of celerite is simplified as: 
\begin{equation}k_{j}(\tau_{nm})=a_{j}e^{-c_{j}\tau_{nm}},
\end{equation}
with the PSD
\begin{equation}
S_j(\omega) = \sqrt{\frac{2}{\pi}} \frac{a_j}{c_j \times \left(1 + \left( \frac{\omega}{c_j} \right)^2 \right)},
\end{equation}
which is referred to as the Damped Random Walk (DRW) or Ornstein–Uhlenbeck process \citep{1930PhRv...36..823U, 1992ApJ...398..169R}. Here $\sigma_{\mathrm{DRW}}$ and $\tau_{\mathrm{DRW}}$ denote the variability amplitude and the characteristic damping timescale, respectively.

The dynamical equation for the SHO term driven by the white noise stochastic driving force $\epsilon(t)$ is given by
\begin{equation}
        \left[ \frac{\mathrm{d}^{2}}{\mathrm{d}t^{2}} + \frac{\omega_{0}}{Q} \frac{\mathrm{d}}{\mathrm{d}t} + \omega_{0}^{2} \right] y(t) = \epsilon(t),
\label{q:5}
\end{equation}
The natural angular frequency is expressed as $\omega_0=2 \pi \nu_0$, where $\nu_0$ is the QPO frequency; and $Q$ is the quality factor. 
The PSD of the SHO term is given by
\begin{equation}
P_{\mathrm{SHO}}(v)=\frac{2S_{0}v_{0}^{4}}{(v^{2}-v_{0}^{2})^{2}+v^{2}v_{0}^{2}/Q^{2}}\ ,
\label{q:3A}
\end{equation}
where $S_0$ is proportional to the power at $\omega=\omega_{0}$, with $S(\omega_0) = \sqrt{\frac{2}{\pi}} \, S_0 Q^2$.
%quantifying the power intensity of periodic or quasi-periodic variations in astronomical time series. 
By matching the coefficients of Eqs.~\ref{q:3} and ~\ref{q:3A} , the parameters $(a_j, b_j, c_j, d_j)$ can be exactly expressed in terms of the SHO parameters $(S_0, \nu_0, Q)$.
The quality factor is defined as $Q = \frac{\omega_0}{4\pi\Gamma}$, where $\Gamma$ denotes the half width at half maximum of the peak in the PSD. The parameter Q quantifies the damping strength: (1) $Q < 0.5$ corresponds to an overdamped regime;
%The system returns to its equilibrium position very slowly without ever crossing it. The strong damping dominates, preventing any oscillates, and causing a sluggish response. 
(2) $Q = 0.5$ represents critical damping;
%the system returns to its equilibrium position in the shortest possible time without crossing it. This corresponds to the boundary between underdamped (oscillatory) and overdamped behavior. 
(3) $Q > 0.5$ corresponds to an underdamped regime, indicating high-quality oscillations. 
The corresponding period $\rho$, damping timescale of the SHO component $\tau_0$, and standard deviation $\sigma$ are given as:
\begin{equation}\rho\equiv2\pi/\omega_{0},\quad\tau_0\equiv2Q/\omega_{0},\quad\mathrm{and}\quad\sigma\equiv\sqrt{S_{0}\omega_{0}Q}.\end{equation}

\subsection{Model fitting process} 
%In the GP framework, the observed data vector $\mathbf{y}$ is modeled as a realization of a stochastic process described by a multivariate Gaussian distribution. A GP can be regarded as an infinite-dimensional generalization of the multivariate normal distribution, fully specified by its the mean function $m$ and $K$. 
Given a set of kernel hyperparameters $\boldsymbol{\phi}$, the likelihood function of the data is written as
\begin{equation}
    p(\mathbf{y} \mid \theta, \phi) = \frac{1}{(2\pi)^{N/2} \vert \mathbf{K} \vert^{1/2}} \exp\left[ -\frac{1}{2} (\mathbf{y} - \mathbf{m})^\top \mathbf{K}^{-1} (\mathbf{y} - \mathbf{m}) \right],
\label{q:1}        
\end{equation}
where $N$ is the number of data points. The covariance matrix is defined as $\mathbf{K} = \mathbf{K}(\boldsymbol{\phi}) + \sigma_n^2 \mathbf{I}$, with $\mathbf{K}(\boldsymbol{\phi})$ representing the kernel that encodes temporal correlations of the data, and $\sigma_n^2 \mathbf{I}$ accounting for observational white noise. 

Bayesian inference is performed by combining the likelihood with prior distributions on the parameters, yielding the posterior distribution
\begin{equation}
p(\theta, \phi \mid y) = \frac{p(y \mid \theta, \phi) p(\theta, \phi)}{p(y)},
\label{q:2}        
\end{equation}
where the evidence $p(y)$ is obtained by marginalizing over the parameter space.

In this work, we use the GP Python package \texttt{celerite2}\footnote{\url{https://celerite2.readthedocs.io/en/latest/index.html}} in combination with Markov Chain Monte Carlo (MCMC) sampling. \texttt{celerite2} kernel functions remain mathematically valid under addition, multiplication, and convolution operations. 
%By combining multiple kernel terms, more complex PSD forms can be constructed \citep{2006gpml.book.....R}. 
The \texttt{celerite2} framework provides an efficient implementation of GP modeling for one-dimensional time series. We use the kernel functions provided by celerite 2 and their additive combinations. To ensure result stability, the MCMC sampler in the program performs 50,000 iterations, with the first 20,000 steps discarded as burn-in. The remaining 30,000 effective samples are used to construct the posterior distributions of the parameters and to reconstruct the GP PSD. During the sampling process, the thinning interval is automatically determined using autocorrelation time analysis to avoid the effects of sample correlation. Two key diagnostic indicators are systematically checked as follows: (1) The standardized residual distribution is approximately matched by the standard normal curve; (2) The residual autocorrelation function (ACF) fluctuates randomly within the $95\%$ white-noise confidence interval. 

%Fig_1-2.pdf
\begin{figure} 
  \includegraphics[width=\linewidth]{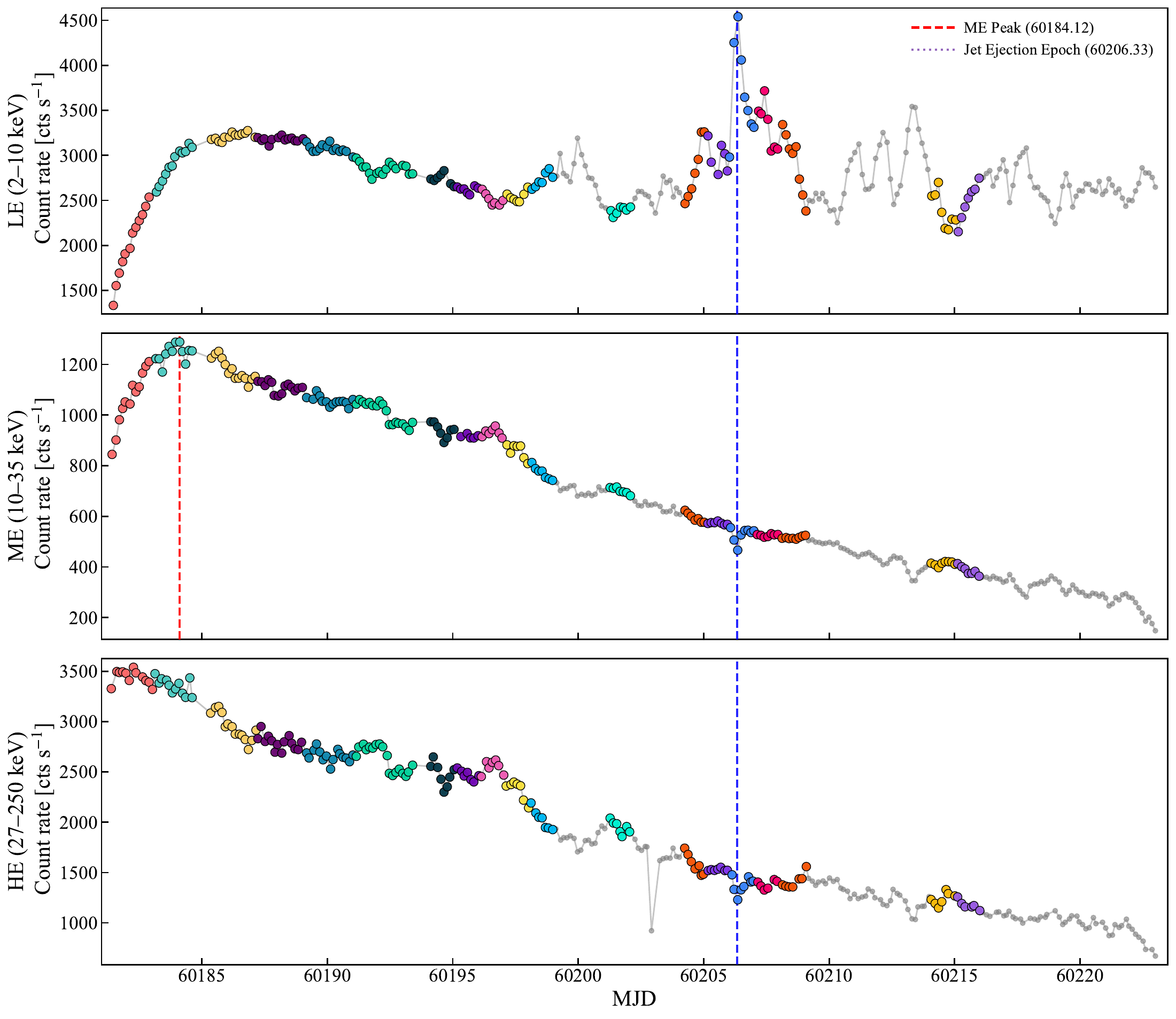}   
      \caption{The light curves of Swift J1727.8–1613 observed by Insight-HXMT. The three panels from top to bottom display the count rates in the LE (2–10 keV), ME (10–35 keV), and HE (27–250 keV) bands, respectively. The gray circles and solid lines illustrate the overall temporal evolution. Specific observations selected for timing analysis are highlighted as colored circles, consistent with the epochs defined in Fig.~\ref{Fig1}. The red vertical dashed line in the middle panel indicates the ME band peak flux observed at MJD 60184.12. A single blue vertical dashed line marks the jet-ejection epoch around MJD~60206.33.}
         \label{Fig2}
\end{figure}

\section{Results}
\label{s:3}
%%%%%%%%%%%%%%%%%%%%%%%%%%%%%%%%%%%%%%%%%%%%%
We selected a time bin size of 0.1 s for all data to capture the temporal structure in the X-ray emissions of Swift J1727.8–1613. A total of 876 GTIs in 19 \textit{ObsIDs} were fitted. The states corresponding to these time intervals are color-coded in Fig.~\ref{Fig1} to illustrate the entire source evolution process. Our primary results are presented in the following section.

\subsection{General GP fitting results}
Previous studies have highlighted the efficacy of composite kernel models \citep{2025MNRAS.537.2380Z,2025A&A...703A.134W}.
\citet{2025A&A...703A.134W} established that, for the black hole X-ray binary MAXI~J1348-630, a model combining SHO, DRW, and AWN components provides accurate characterization of its X-ray light curves. They further proposed that the AWN component should be interpreted as a rapid DRW fluctuation. 
This finding provides a critical informative prior for our current modeling of Swift J1727.8–1613. 
We therefore adopt a SHO+DRW+DRW composite kernel. 
We have seven free hyper-parameters: $S_0$, $\nu_0$, and $Q$ for the SHO component; 
$\sigma_{\rm long}^{\rm DRW}$ and $\tau_{\rm long}^{\rm DRW}$ for the long-timescale DRW component; 
and $\sigma_{\rm short}^{\rm DRW}$ and $\tau_{\rm short}^{\rm DRW}$ for the short-timescale DRW component. 
In this model, the SHO component directly captures the QPO signal in the data, 
and the two DRW components model the broadband noise, with the long-timescale DRW tracing low-frequency fluctuations and the short-timescale DRW tracing high-frequency variability.

%HE.pdf
\begin{figure*} 
  \centering
  \begin{subfigure}[b]{0.48\linewidth}
    \includegraphics[width=\linewidth]{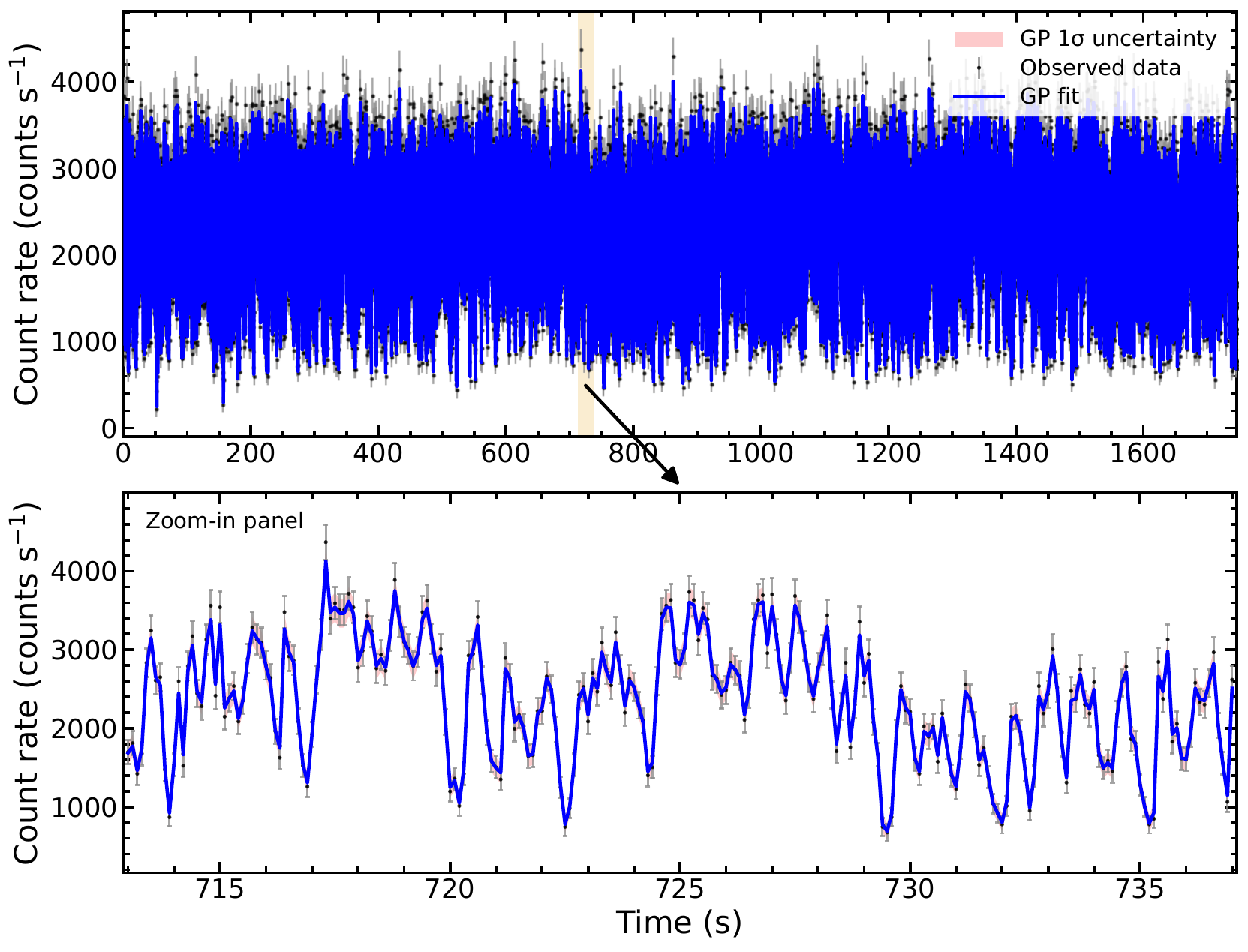}
    \label{Fig1a}
  \end{subfigure}
  \hfill
  \begin{subfigure}[b]{0.48\linewidth}
    \includegraphics[width=\linewidth]{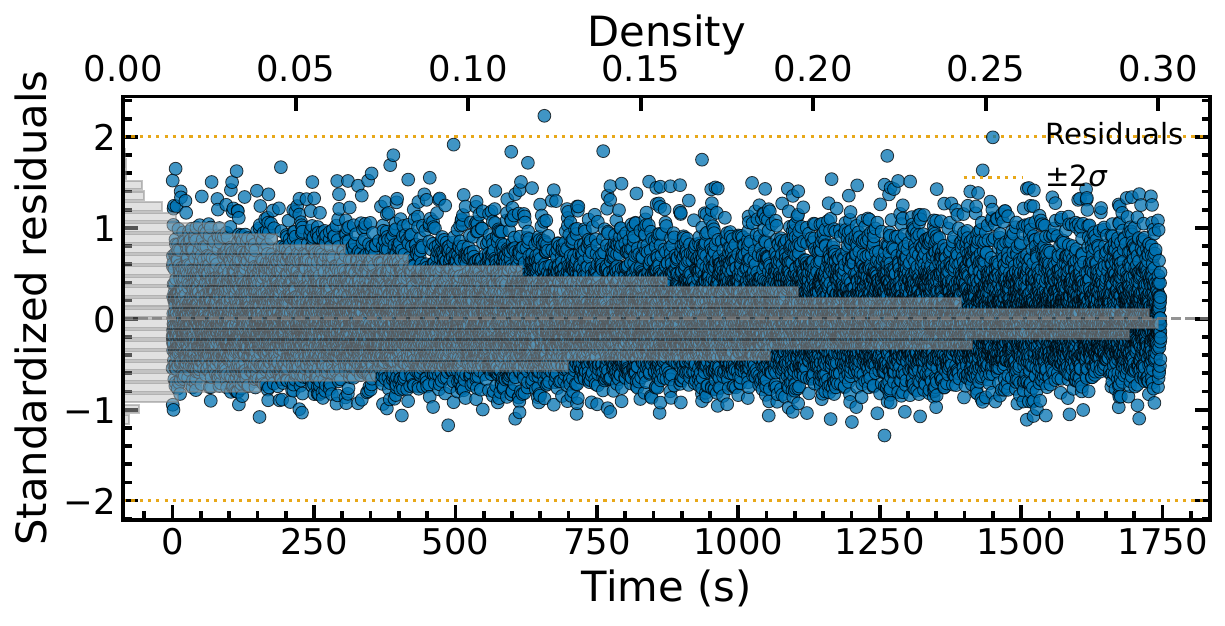}
    \label{Fig1b}
  \end{subfigure}
  
  \vspace{0.5em} % 可选的垂直间距
  
  \begin{subfigure}[b]{0.48\linewidth}
    \includegraphics[width=\linewidth]{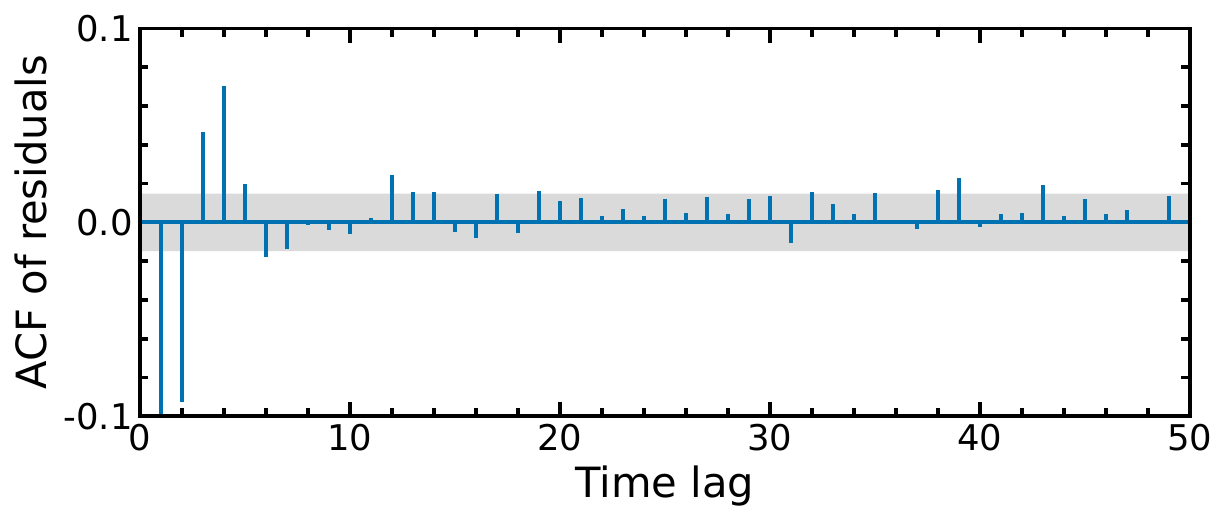}
    \label{Fig1c}
  \end{subfigure}
  \hfill
  \begin{subfigure}[b]{0.48\linewidth}
    \includegraphics[width=\linewidth]{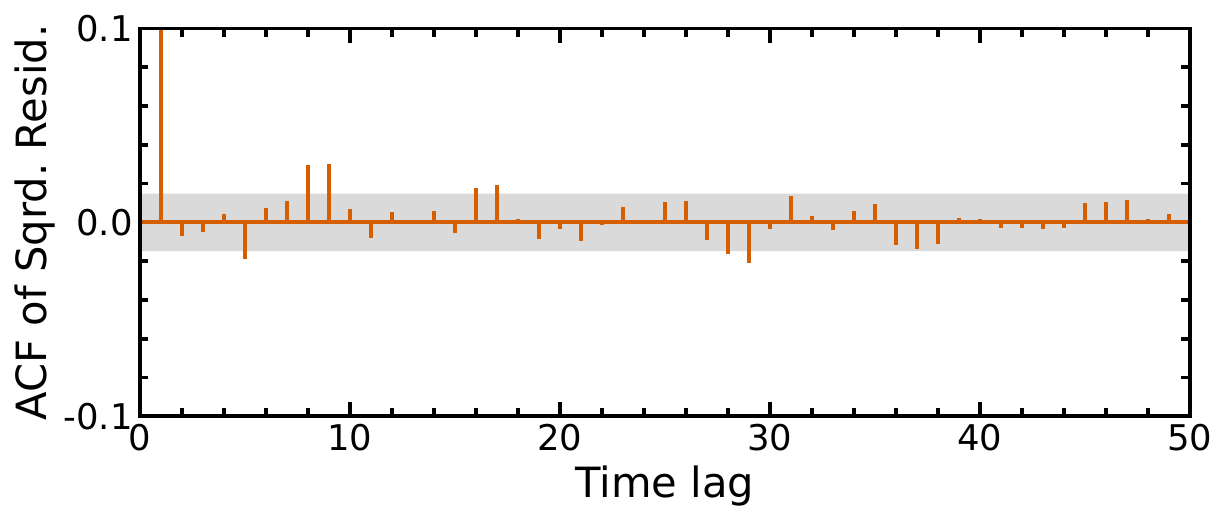}
    \label{Fig1d}
  \end{subfigure}
  
  \vspace{0.5em} % 可选的垂直间距
  
  \begin{subfigure}[b]{0.48\linewidth}
    \includegraphics[width=\linewidth]{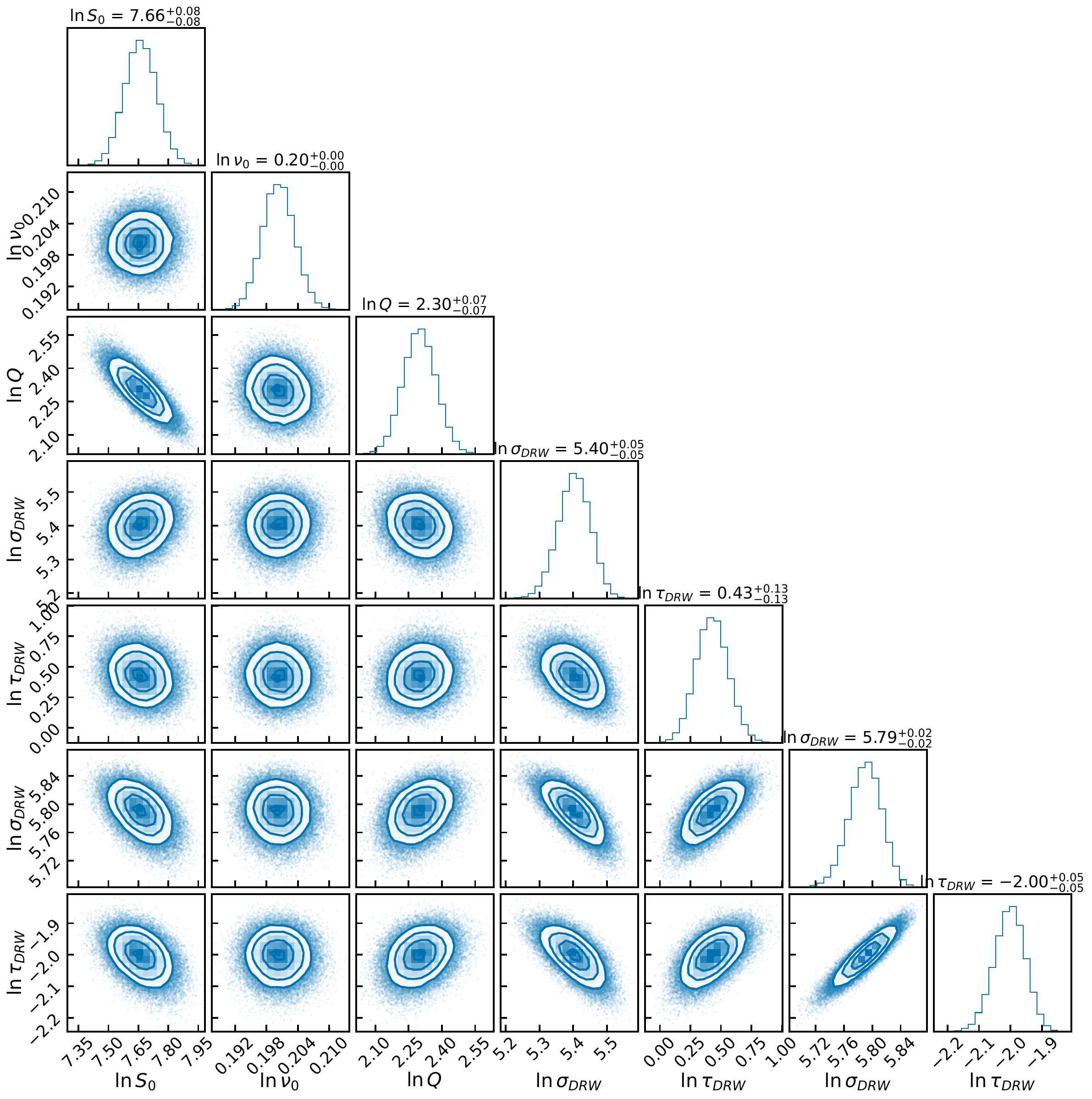}
    \label{Fig1e}
  \end{subfigure}
  \hfill
  \begin{subfigure}[b]{0.48\linewidth}
    \includegraphics[width=\linewidth]{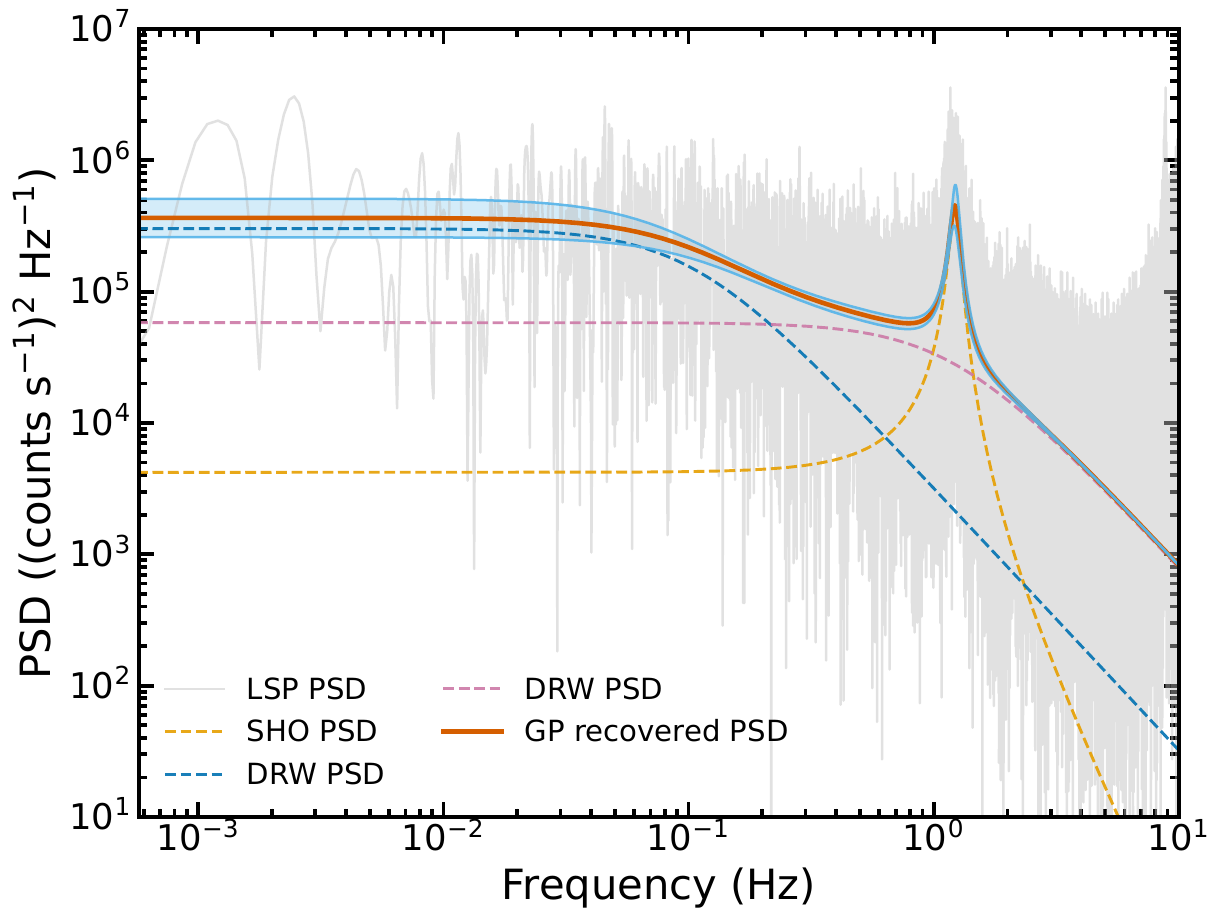}
    \label{Fig1f}
  \end{subfigure}
  
  \caption{
  \textbf{Top left}: The HE light curve and its GP fit. The main panel shows the observed data as black points, the best-fit GP model as a blue line, and the 1$\sigma$ uncertainty as the shaded region. The yellow highlighted interval is enlarged in the zoom-in panel below, which shows the detailed light-curve variation and GP fit over the selected time interval.
  \textbf{Top right}: Standardized residuals and their corresponding density distribution.
  \textbf{Middle row}: The ACFs of the residuals left and squared residuals right.
  \textbf{Bottom left}: Corner plot showing the posterior probability distributions of the GP hyperparameters.
  \textbf{Bottom right}: PSD decomposition. The total GP-recovered PSD red is compared with the Lomb--Scargle periodogram gray. The individual components include one SHO kernel yellow dashed and two DRW kernels blue and pink dashed.}
  \label{Fig3}
\end{figure*}

Fig.~\ref{Fig3} presents the comprehensive results of the GP modeling applied to the observed light curve (the example of HE data of ObsID P061433800602). It illustrates the time-domain fit, the spectral decomposition of the kernel components, the posterior parameter distributions, and a systematic analysis of the residuals.
The top left panel shows the observed light curve (black) overlaid with the GP predictive mean (blue) and the $1\sigma$ uncertainty region (red). 
The model visually traces the variability trends across both long and short timescales.
The GP PSD plot (bottom right) demonstrates that the GP model (red solid line and blue shadow) successfully recovers the spectral features of the data. The model decomposes the variability into two red noise components modeled by two different DRWs (dashed lines) and a significant QPO modeled by the SHO kernel with $\nu_0\approx 1.2$ Hz.
The corner plot (bottom left) displays the posterior distributions of the kernel hyperparameters obtained via MCMC sampling. The quasi-Gaussian profiles in the 1D histograms and the well-constrained contours in the 2D planes indicate good convergence of the chain and that all parameters are well-determined.

To evaluate the accuracy of the GP model, 
we performed a systematic diagnostic on the residuals obtained after 
subtracting the predictive mean of the SHO+DRW+DRW structure from the observed light curve. Under the assumption of an ideal fit, the residuals should approximate independent and identically distributed Gaussian white noise.  
The distribution of residuals (top right panel) appears stationary in time, 
with no evidence of significant variance evolution or volatility clustering.
The ACF of the squared residuals (middle right panel) shows no significant correlation, remaining well within the 95\% confidence intervals. 
%This indicates the absence of autoregressive conditional heteroscedasticity (ARCH) effects or volatility clustering. 
It confirms that the variance of the noise is correctly modeled and stationary.
%, supporting the assumption of homoscedasticity inherent in the standard GP formulation.
The distribution of standardized residuals approximates a Gaussian but exhibits a slight positive skewness. This deviation is likely attributable to the intrinsic log-normal flux distribution characteristic of accretion-powered sources, i.e., the linear rms-flux relation. Since the GP modeling was performed in the linear count rate space, this multiplicative nature of the variability introduces a minor asymmetry in the additive residuals. Given that the deviation is small and the spectral recovery remains robust, we proceed with the current linear-space model.

However, the ACF of the residuals reveals significant structured correlation at short time lags (middle left panel). 
Specifically, the ACF exhibits a statistically significant oscillatory pattern within the first few lags ($t \lesssim 0.5$\,s), with amplitudes exceeding the 95\% white-noise confidence intervals. The characteristic {\it negative-to-positive} fluctuation 
at lags 1-4 implies the presence of a coherent signal with the frequency of $\nu\approx1/0.4\rm{s}=2.5\ $Hz that was not captured by the SHO term.
The LSP PSD indeed shows an excess power localized at $\nu_{\mathrm{res}} \approx 2.5$\,Hz. Notably, this frequency stands in an integer ratio to the fundamental QPO frequency extracted by our model ($\nu\approx 2\nu_0$). This relationship suggests that the residual feature corresponds to the second harmonic of the primary oscillation. The presence of such harmonic content indicates that the underlying QPO waveform is intrinsically non-sinusoidal.

\subsection{Evolution of the SHO component parameters} 
%SHO
\begin{figure*} 
  \includegraphics[width=\linewidth]{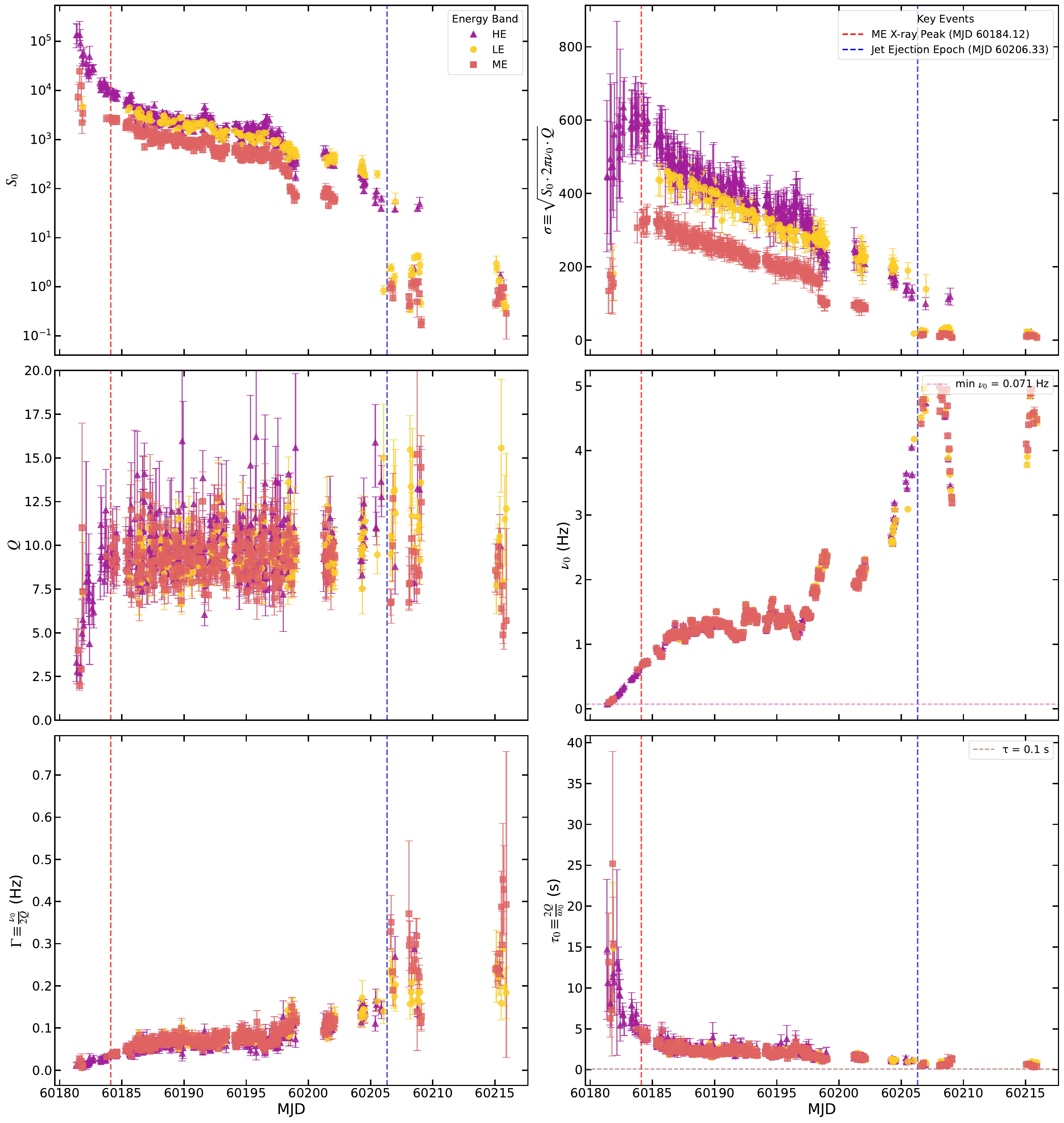}   
      \caption{Temporal evolution of the SHO model parameters. Different markers distinguish the energy bands: HE (magenta purple triangles), ME (coral red squares), and LE (yellow circles). The red vertical dashed line marks the peak of the ME X-ray flux. A single blue vertical dashed line marks the jet-ejection epoch around MJD~60206.33.}
         \label{Fig4}
\end{figure*}
 
Fig.~\ref{Fig4} presents the evolution of the SHO hyperparameters derived from the HE, ME and LE light curves. It shows the temporal variations of $S_0$, $\nu_0$, $Q$, and the corresponding changes in $\sigma$, $\Gamma$, and $\tau_0$. The vertical dashed lines denote critical timing markers, including the ME X-ray peak and the jet-ejection events.

%%$S_0$ $\sigma$
The upper panels display the temporal evolution of $S_0$ (left) and $\sigma$ (right). 
In the top-left panel, $S_0$ is plotted on a logarithmic scale to capture its systematic reduction from $\sim 10^{5}$ to $\sim 0.1$ during the outburst. 
A sharp transition occurs immediately following the jet ejection events at MJD~60206, resulting in a drop of more than two orders of magnitude across all bands. Coincidentally, after MJD~60205, when the flaring activity begins, the QPO is observed to transition from type-C to type-B, with the frequency reaching up to $\sim10~\mathrm{Hz}$ \citep{2024MNRAS.531.4893M}.
In the top-right panel, the value of $\sigma$ increases rapidly to a peak of $\sim 600$ before MJD 60184 and subsequently decays toward a minimum level of $\sim 1$ in the late phase.  
%Both parameters characterize the signal strength and exhibit a clear energy dependence, with the HE and LE bands consistently showing higher values than the ME band.  Notably, in the LE band, both $S_0$ and $\sigma$ exceed the levels of the HE band around MJD 60198, subsequently declining in unison with the HE and ME bands.

%%Q
The middle-left panel shows the evolution of $Q$. 
The $Q$ initially increases from $\sim 3$ to $\sim 10$ prior to the peak of the ME-band X-ray intensity at MJD~60184.12. 
After MJD~60184.12, $Q$ remains stable and fluctuates around a value of $\sim 10$. This transition occurs slightly earlier than the peak of the onset of the outburst in the LE band, but later than the peak of the outburst in the HE band, as shown in Fig.~\ref{Fig2}.

%%$\nu_0$
The middle-right panel shows the temporal evolution of $\nu_{0}$. At MJD 60182, $\nu_{0}$ is measured to be $\sim0.07~\mathrm{Hz}$. Then it increases rapidly and reaches $\sim1.3~\mathrm{Hz}$ by MJD 60187. This rapid increase is consistent with previous studies \citep{2023ATel16235....1K, 2023ATel16243....1K, 2024MNRAS.531.4893M}. Between MJD 60187 and 60190, $\nu_{0}$ remains constant at $\sim1.3~\mathrm{Hz}$. 
A similar plateau phase was reported by \cite{2025ApJ...993...40X}, 
although in their analysis $\nu_{0}$ persisted at $\sim1.3~\mathrm{Hz}$ until MJD 60197. 
After MJD 60190, $\nu_0$ exhibits several rapidly evolving peaks. In the systematic analysis with $\Delta t=0.1$~s, the recovered frequency reaches approximately 5 Hz. We reanalyzed selected LE light curves near MJD 60206 using $\Delta t=0.01$~s, while retaining the same SHO+DRW+DRW model and fitting procedure. The higher-resolution analysis recovers a rapid increase in $\nu_0$ from approximately 4 to 9 Hz over about eight hours, followed by a decline to approximately 5 Hz. This evolution is consistent with the results reported by \cite{2024MNRAS.531.4893M, 2025ApJ...993...40X}.

%%$\Gamma$  $\tau_0$
The bottom panels show the $\Gamma$ (left) and $\tau_0$ (right). The value of $\Gamma$ increases gradually from $\sim 0.01$ to $\sim 0.5 \mathrm{~Hz}$ over the duration of the observation. Conversely, the $\tau_0$ exhibits a sharp reduction. Between MJD 60182 and 60185, $\tau_0$ decreases from $\sim 25 \mathrm{~s}$ to  $\sim4 \mathrm{~s}$. Following this initial drop, the timescale continues a more gradual decline, eventually approaching the $0.1 \mathrm{~s}$ of sampling limit indicated by the dashed horizontal line.

\subsection{Evolution of the two DRW component parameters} 
%DRW
\begin{figure*} 
  \includegraphics[width=\linewidth]{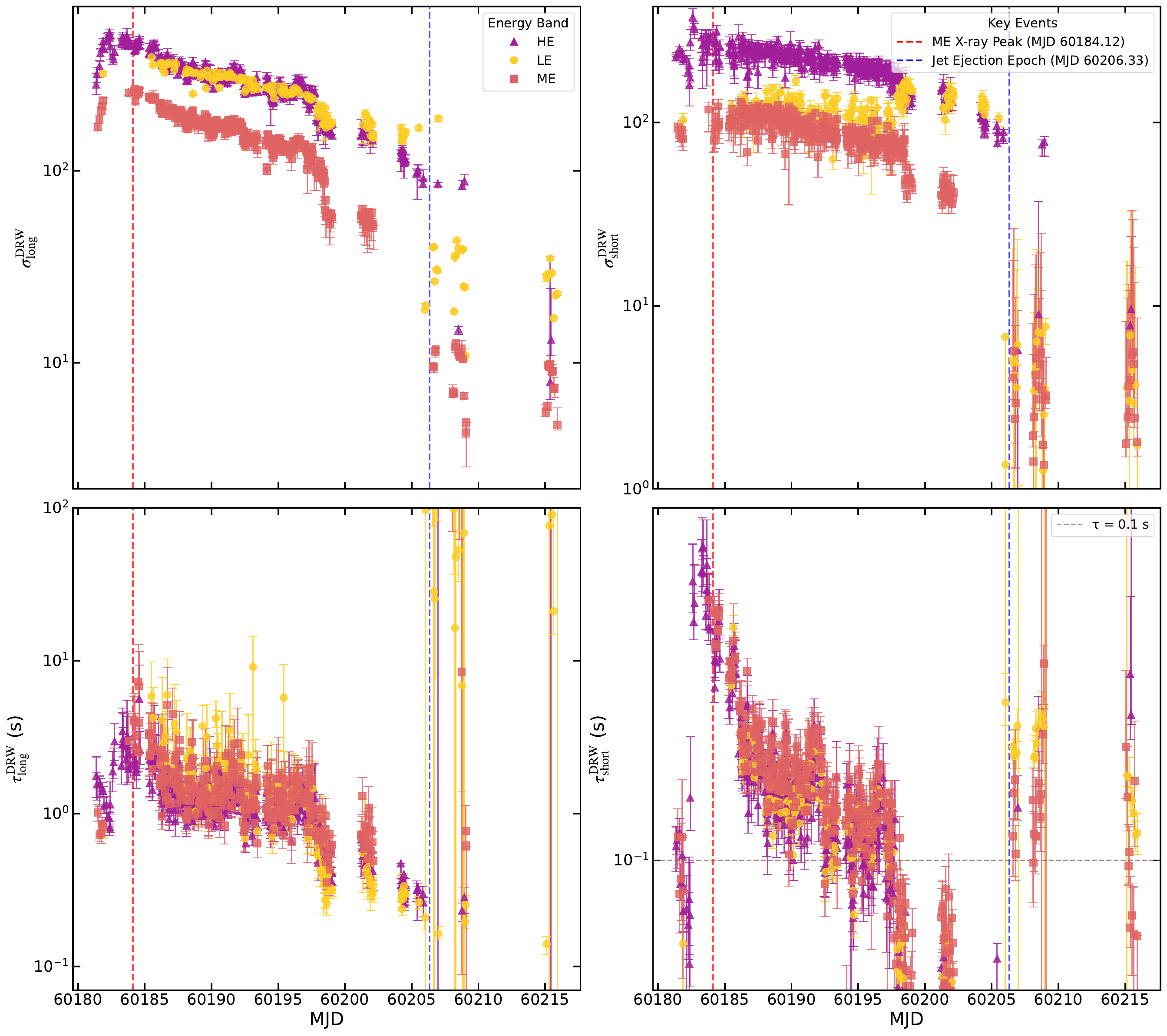}   
      \caption{Temporal evolution of the two DRW model parameters. Same as Fig.~\ref{Fig4}. }
         \label{Fig5}
\end{figure*}

Fig.~\ref{Fig5} presents the temporal evolution of the two DRW hyperparameters. 
The upper panels display the evolution of long-timescale amplitude DRW component $\sigma_{\mathrm{long}}^{\mathrm{DRW}}$ (left) and short-timescale amplitude DRW component $\sigma_{\mathrm{short}}^{\mathrm{DRW}}$ (right). Throughout the outburst, both $\sigma_{\mathrm{long}}^{\mathrm{DRW}}$ and $\sigma_{\mathrm{short}}^{\mathrm{DRW}}$ exhibit a sustained declining trend across all energy bands. In the $\sigma_{\mathrm{long}}^{\mathrm{DRW}}$ panel, the HE and LE bands consistently maintain higher magnitudes compared to the ME band. From MJD~60182 to 60184.12, $\sigma_{\mathrm{long}}^{\mathrm{DRW}}$ in the HE and LE bands increases from $\sim 200$ to $\sim 500$, while the ME band rises from $\sim 170$ to $\sim 250$. Between MJD~60184.12 and 60198, these amplitudes decrease gradually to approximately 200 (HE/LE) and 100 (ME). During the late phase (MJD~60198-60216), the amplitudes further decay to $\sim 15$ for the HE/LE bands and $\sim 3$ for the ME band. 

In the $\sigma_{\mathrm{short}}^{\mathrm{DRW}}$ panel, the HE band initially dominates with an amplitude of $\sim 300$ before declining to 100 by MJD~60198. Meanwhile, the ME and LE bands remain at a lower baseline of $\sim 100$ and gradually decrease to 70. During MJD 60198-60216, the amplitudes in all bands undergo a rapid decay, reaching values of order unity. A notable feature occurs around MJD~60197, where both $\sigma$ parameters in the LE band experience a transient increase, briefly exceeding the HE band before returning to the global decay. 
%Following the jet ejection events at MJD~60206 (marked by vertical dashed lines), both DRW amplitudes drop abruptly by more than one order of magnitude across all energy bands.

The bottom panels show the temporal evolution of the long-timescale DRW component $\tau_{\mathrm{long}}^{\mathrm{DRW}}$ (left) and the short-timescale DRW component $\tau_{\mathrm{short}}^{\mathrm{DRW}}$ (right). The $\tau_{\mathrm{long}}^{\mathrm{DRW}}$ increases from $\sim 0.7~\mathrm{s}$ to $\sim 6~\mathrm{s}$ between MJD~60180 and 60185, followed by a stable phase at $\sim 1~\mathrm{s}$ until MJD~60197. Subsequently, it decreases toward 0.1~s leading up to MJD~60206.
%Following the ejection, $\tau_{\mathrm{long}}^{\mathrm{DRW}}$ becomes poorly constrained, because the two DRW components are strongly degenerate and effectively merge into a single DRW term.
%consistent with the unconstrained DRW parameters reported by \cite{2025A&A...703A.134W} for the HE and ME energy bands.
The $\tau_{\mathrm{short}}^{\mathrm{DRW}}$ increases from 0.1 s at MJD 60180 to 0.4 s at MJD 60184, before decreasing to 0.2 s by MJD 60186. During MJD 60186-60197, it further declines from 0.2 s to 0.1 s. In the subsequent interval (MJD 60197–60206), the short timescale remains consistently below 0.1 s. After MJD 60206, it fluctuates between 0.1 s and 0.2 s. In the case of $\tau_{\mathrm{short}}^{\mathrm{DRW}}\approx0.1\ $s, 
the fast DRW effectively reduces to a proxy for white-noise–like fluctuations in the observed variability.

Following the jet ejection events at MJD 60206, both DRW amplitudes drop abruptly by more than one order of magnitude across all energy bands. During this late phase, as the parameters of the two kernels become comparable, the two DRW components become highly degenerate and effectively merge into a single red-noise process.

It is worth noting that during the fitting of individual GTIs, the amplitude of the short-timescale DRW component ($\sigma_{\rm short}^{\rm DRW}$) sometimes becomes extremely small, rendering its contribution to the overall variability practically negligible. Interestingly, this phenomenon exhibits a statistical dependence on both the energy band and the spectral state. Specifically, during the hard state (e.g., prior to MJD 60190), a substantial proportion of the GTIs in the LE and ME bands do not require this short-timescale DRW component, although it remains necessary for a certain fraction of the intervals. In contrast, the fits to the HE band light curves during the same epochs rarely exhibit this absence, with the short-timescale DRW component remaining prominent in almost all GTIs. A detailed statistical validation of this fractional model degeneracy via Bayesian model selection is beyond the scope of this paper and will be systematically investigated in a subsequent work.

Because the light curves have a sampling interval of 0.1 s, DRW damping timescales at or below 0.1 s are not treated as resolved measurements. At these timescales, most of the temporal covariance decays within one time bin, and the DRW component approaches a white-noise-like limit. The two DRW components can also become strongly degenerate. We therefore exclude these measurements from Fig.~\ref{Fig9} and ~\ref{Correlations} and from the quantitative correlation and viscosity analyses.

\section{A Stochastic Magneto-Active Dynamics Scenario}\label{sect_discussion}
\label{s:4}
%%%%%%%%%%%%%%%%%%%%%%%%%%%%%%%%%%%%%%%%%%%%%
We establish a correspondence between the statistical modeling of observed variability and underlying physical dynamics by formulating the light curve as the realization of a linear SDE driven by white noise. In this dynamical framework, the system’s response to stochastic excitation is encoded in its Green's function $G(t)$. Crucially, 
the covariance structure used in our GP regression is not an arbitrary choice, 
but is physically constrained to be the autocorrelation of this impulse response:
$$k(\tau) = \int G(t)G(t+\tau) dt$$
This formalism allows us to transcend light curve fitting. 
Consequently, the kernel hyperparameters provide statistical measures that can be compared with characteristic physical quantities, permitting the precise extraction of dynamical constraints like the damping timescale $\tau_{\rm damping}$ and the damping ratio $\zeta=1/2Q$ from the light curves.
%This framework maps GP hyperparameters onto SDE coefficients, enabling the retrieval of the underlying dynamical process and its key parameters, including $\tau$ and $\zeta=1/2Q$, from observational data.

\subsection{Stochastic Magneto-Active Dynamics}
Guided by our GP results, we propose a stochastic magneto-active dynamics scenario to understand the physics under the three components. We conceptualize the inner accretion flow around a black hole as a magneto–inertial resonant cavity confined by an ordered, large-scale magnetic field. Its geometry and dynamical stability are jointly regulated by the gravitational potential, plasma thermal pressure, and magnetic tension. In this picture, curved coherent field lines act as a {\it magnetic spring}, providing the dominant restoring force against perturbations, whereas dense, field-aligned plasma provides the effective inertia. Stochastic driving by turbulence and magnetic reconnection excites the system, producing quasi-resonant oscillations near its eigen-frequencies; dissipative processes then remove energy from these modes, yielding damped responses in the observed variability. This magnetic-cavity picture is similar to the oscillating hot-corona model of \cite{2010MNRAS.404..738C}.

The eigenfrequency $\omega_0$ is determined by the Alfv\'en time scale and the cavity size $L$.
It is defined as
\begin{equation}
\omega_0 = k_{||} v_A \approx \frac{\pi v_A}{L}
\end{equation}
where $k_{\parallel}$ denotes the component of the wave vector parallel to the large-scale magnetic field $B_0$, and $v_{\rm A}=B_0/\sqrt{4\pi\rho_{\rm ion}}$ is the Alfv\'en speed; $\rho_{\rm ion}$ denotes the mass density dominated by ions.

%%%%%%%Q
The quality factor quantifies the coherence of the oscillation mode and is defined as the ratio of the natural angular frequency to twice the total amplitude damping rate \citep[e.g.,][]{1945RvMP...17..323W}, 
\begin{equation}
Q = \omega_0/(2\gamma_{\mathrm{total}}).
\label{Q:11}
\end{equation}
This relation follows from the standard damped-harmonic oscillator equation.
Here, $\gamma_{\mathrm{total}}$ denotes the rate of amplitude damping and, in our phenomenological scenario, includes dissipation through turbulent cascades and energy leakage through the boundaries of the cavity.

In this framework, the MHD-driven dynamical perturbations are not observed directly but imprint themselves on the X-ray emission by modulating the disk-coronal thermodynamic state. 
Specifically, fluctuations associated with the global oscillatory mode and the fast/slow stochastic relaxation channels perturb the electron temperature, optical depth, and the seed-photon field. Under the small-perturbation approximation and assuming rapid radiative adjustment relative to the dynamical timescales, 
the radiative transfer response can be linearized. 
Therefore, the flux variation in each energy band can be written as a weighted sum of the three underlying variability components: the global oscillatory mode and the fast and slow stochastic relaxation processes. The weight assigned to each component quantifies how strongly that energy band responds to the corresponding physical process. 
This provides a direct statistical projection from MHD variability into multi-band X-ray light curves and their covariance structure.

\begin{figure} 
  \includegraphics[width=\linewidth]{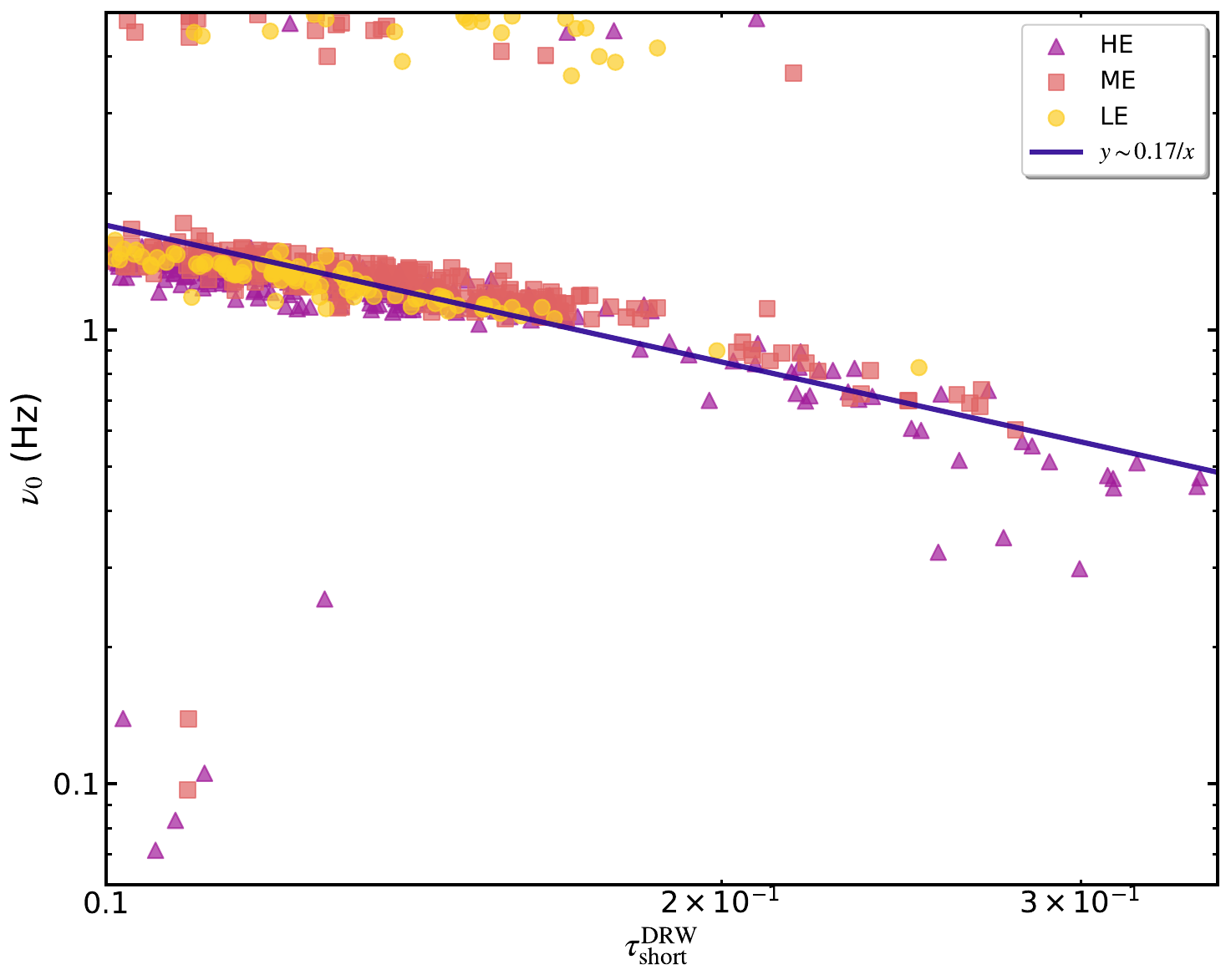}   
      \caption{Relations between $\nu_0$ and $\tau_{\mathrm{short}}^{\mathrm{DRW}}$ in the HE (magenta purple triangles), ME (coral red squares), and LE (yellow circles) energy bands.
    }
         \label{Fig9}
\end{figure}

In this scenario, the following relation is directly predicted:
\begin{equation}
\omega_0 \propto \frac{v_A}{L} \propto \frac{1}{\tau^{\rm DRW}_\mathrm{short}}.
\end{equation}
%This relation is consistent with the result shown in Fig.~\ref{Fig9}, which illustrates the correlation between the QPO frequency $\nu_0$ of the SHO component and the short timescale $\tau_\mathrm{short}^\mathrm{DRW}$ of the DRW component. Specifically, $\nu_0$ is found to be anti-correlated with $\tau_\mathrm{short}$.

\subsection{Linking the Physical Model to the GP results} 
%$\nu_0$
Within the framework of stochastic magneto-active dynamics, we consider a progressive gravitational contraction of the disk-coronal resonant cavity, such that its characteristic size $L$ decreases with time. 
Allowing the effective mass participating in the oscillation, $M_{\mathrm{cav}}$, to evolve with time, the mean density scales as $\rho \propto M_{\mathrm{cav}} L^{-3}$. Under magnetic flux conservation, the mean magnetic field strength scales as $B_0 \propto L^{-2}$, which gives $v_{\mathrm{A}} \propto L^{-1/2} M_{\mathrm{cav}}^{-1/2}$ and $\nu_0 \propto L^{-3/2} M_{\mathrm{cav}}^{-1/2}$. Thus, for comparable fractional changes, the dependence of $\nu_0$ on $L$ is stronger than that on $M_{\mathrm{cav}}$.

Between MJD 60182 and 60187, the rapid rise of $\nu_0$ indicates an initial contraction of the disk-coronal resonant cavity, corresponding to a decrease in $L$.
From MJD 60187 to 60190, $\nu_0$ remains constant at  $\sim1.3$~Hz, suggesting that the system temporarily attains a steady configuration. During this interval, the gravitational potential, plasma thermal pressure, and magnetic tension appear to be mutually balanced, resulting in a relatively stable cavity size.
During MJD 60190-60207, $\nu_0$ exhibits pronounced variability with intermittent peaks reaching up to $\sim 5$~Hz, indicating a phase of continued cavity contraction. 
This progressive reduction of $L$ concentrates magnetic flux within an increasingly compact region, driving the gravitational, thermal, and magnetic energies toward its stability limit.
Between MJD 60207 and 60210, $\nu_0$ drops abruptly from $\sim 5$~Hz to $\sim 3$~Hz, corresponding to an expansion of the resonant cavity. 
During MJD 60210–60216, the $\nu_0$ recovery to $\sim 5$~Hz indicates secondary compression, re-establishing a compact disk–corona configuration.

%$ Q $ 
Between MJD 60182 and 60184, $Q$ rises from $\sim3$ to $\sim10$, while $\nu_0$ increases more rapidly. The increase of $Q$ is slower than linear with $\nu_0$, indicating that $\gamma_{\rm total}$ also increases over this interval (Eq.~\ref{Q:11}). Although the contraction of the disk-coronal cavity drives a rapid increase in $\nu_0$, the concurrent growth of $\gamma_{\rm total}$ partially suppresses the increase of $Q$.
After MJD 60184, $\nu_0$ continues to increase while $Q$ remains roughly constant at $\sim10$. This suggests that $\gamma_{\rm total}$ increases approximately in proportion to $\nu_0$, maintaining an approximate balance between the numerator and denominator in Eq.~(\ref{Q:11}).
Consequently, the system sustains a highly coherent oscillation with a stable $Q \sim 10$ (see Fig.\ref{Fig4}).

%$two DRW$
%The SHO period $P_{\rm SHO}$ decreases from 14~s to 0.2~s, reflecting a decrease in the physical size of the resonant cavity.  
The long DRW timescale approaches the 0.1 s sampling limit near MJD 60206, while the short timescale is already unresolved. Their apparent convergence may result from the sampling limit. The data nevertheless indicate increasingly rapid variability near this stage, and higher-cadence observations will be required to determine the evolution below 0.1 s.

The pronounced attenuation in the amplitudes of both the SHO component ($S_0$) and the stochastic DRW components ($\sigma$), characterized by an abrupt and dramatic collapse immediately preceding the jet ejection at MJD~60206, signifies a fundamental redistribution of the system's energy budget. This precipitous decline suggests that the power previously fueling internal magneto-thermal fluctuations and coherent Alfvén oscillations within the disk-coronal environment is channeled into the kinetic energy required to drive the relativistic outflow. 

Fig.~\ref{Fig9} shows the relationship between $v_{0}$ and $\tau_{\rm short}^{\rm DRW}$ across all energy bands. We find a clear anti-correlation, where the increase in oscillation frequency is accompanied by a systematic decrease in the fast stochastic relaxation time. This trend is consistent with our proposed stochastic magneto-active dynamics, in which contraction of the resonant cavity simultaneously increases the eigenfrequency and shortens the local turbulent turnover time. 
%Notably, at high frequencies ($v_{0} > 3$~Hz), the $\tau_{\rm short}^{\rm DRW}$ values cluster near the 0.1~s dashed line, representing the sampling limit of our light curves. This clustering suggests that while the dynamical processes may continue to accelerate, their precise characterization becomes constrained by the instrumental time resolution.

\subsection{
Possible relation to the effective viscosity parameter
}
\label{subsec:alpha_estimator} 

The GP decomposition characterizes the statistical structure of the observed X-ray variability and provides characteristic stochastic timescales. These timescales themselves do not directly correspond to specific physical processes. Instead, their possible physical meanings are interpreted by comparing their evolution and relative hierarchy with characteristic timescales expected in accretion-flow physics.
 
The short-timescale DRW component describes the higher-frequency broadband variability, and its damping timescale is anti-correlated with the QPO frequency, as shown in Fig.~\ref{Fig9}. This behavior is consistent with local MHD turbulence and magnetic reconnection becoming faster as the characteristic size of the disk--coronal region decreases. In magnetorotationally turbulent accretion flows, local turbulent fluctuations evolve on timescales comparable, within an order-unity factor, to the local orbital or dynamical timescale \citep{1995ApJ...440..742H, 1996ApJ...463..656S}. This motivates the association $\tau_{\mathrm{short}}^{\mathrm{DRW}} \sim t_{\mathrm{dyn}}$.

The long-timescale DRW component describes a slower, lower-frequency relaxation channel and is generally approximately one order of magnitude longer than the short-timescale component during the resolved part of the outburst. We therefore consider it a plausible tracer of the thermal adjustment of the accretion flow. In the standard $\alpha$ prescription, the thermal timescale satisfies $t_{\mathrm{th}} \sim t_{\mathrm{dyn}}/\alpha$ \citep{1973A&A....24..337S}, motivating the association $\tau_{\mathrm{long}}^{\mathrm{DRW}} \sim t_{\mathrm{th}}$.

\begin{figure} 
  \includegraphics[width=\linewidth]{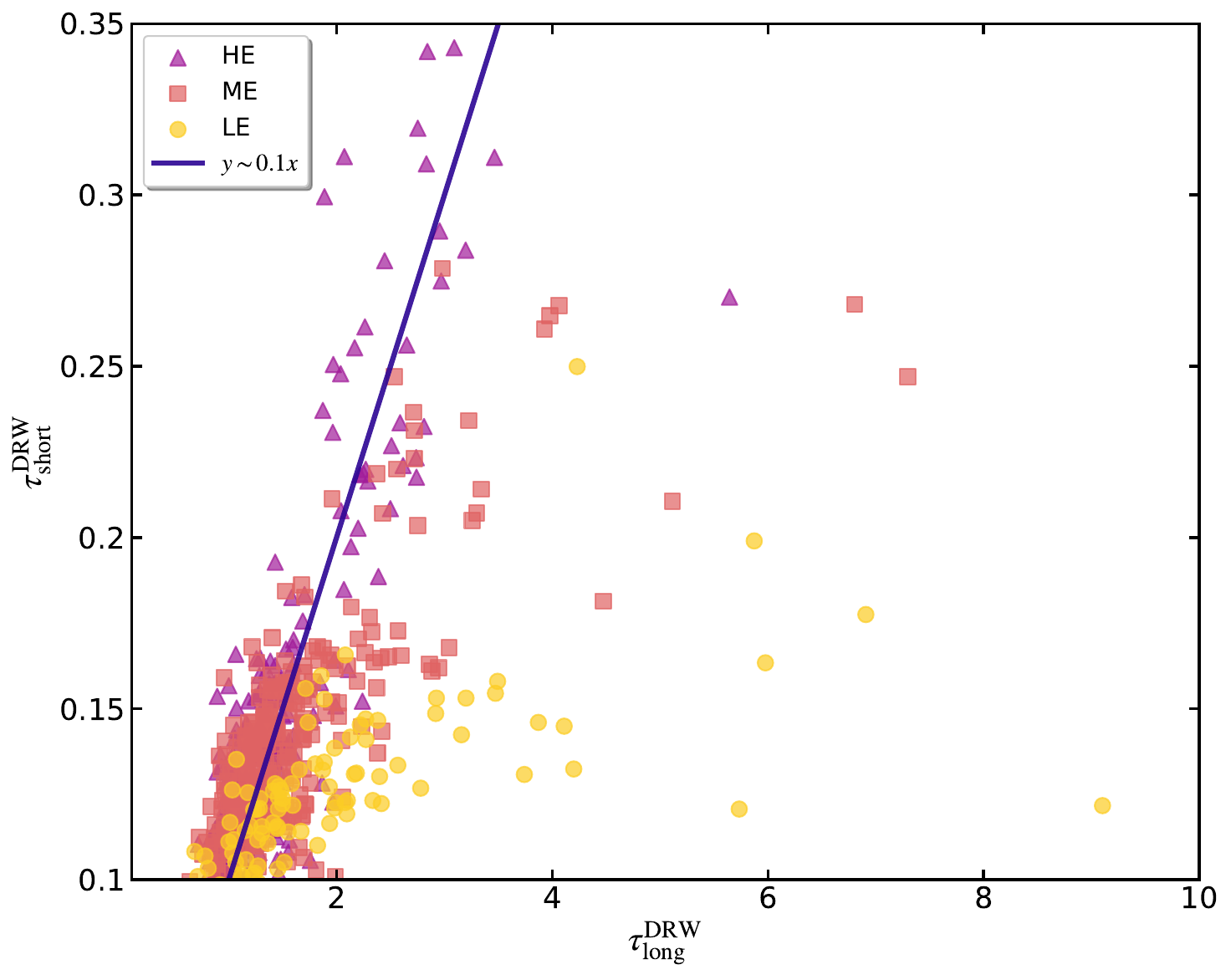}   
      \caption{Correlations between $\tau_{\mathrm{long}}^{\mathrm{DRW}}$ and $\tau_{\mathrm{short}}^{\mathrm{DRW}}$. The color coding matches Fig.~\ref{Fig9}. The solid line shows the reference relation in which the short timescale is one tenth of the long timescale. It is not a formal fit to the data.
    }
         \label{Correlations}
\end{figure}

By relating the stochastic damping timescales derived from our GP modeling to the characteristic timescales of the accretion disk, we examine whether the observed timescale ratio is consistent with the expected scaling of the effective viscosity parameter. Assuming that $\tau_{\mathrm{short}}^{\mathrm{DRW}}$ traces the turbulent turnover timescale ($t_{\mathrm{turb}} \sim t_{\mathrm{dyn}}$), and that $\tau_{\mathrm{long}}^{\mathrm{DRW}}$ traces the thermal timescale ($t_{\mathrm{th}}$), the standard relation $t_{\mathrm{th}} \sim t_{\mathrm{dyn}}/\alpha$ implies $\alpha_{\mathrm{obs}} \sim \tau_{\mathrm{short}}^{\mathrm{DRW}} / \tau_{\mathrm{long}}^{\mathrm{DRW}}$.

Fig.~\ref{Correlations} shows the relation between the long and short stochastic timescales obtained from the GP decomposition. During most of the outburst, particularly in the HIMS, the two timescales follow an approximately linear relation, with their ratio generally clustering around $\tau_{\rm short}^{\rm DRW}/\tau_{\rm long}^{\rm DRW}\sim0.1$. Under the adopted physical interpretation, the resolved DRW timescale ratio is therefore consistent with an effective viscosity parameter of order $\alpha \sim 0.1$.

Special attention is directed toward the behavior near MJD 60206, coincident with the onset of the relativistic jet ejection. During this phase, both $\tau_{\rm long}^{\rm DRW}$ and $\tau_{\rm short}^{\rm DRW}$ decrease dramatically, with the data points moving toward the lower-left region of the plot. The long DRW timescale approaches the 0.1 s sampling limit, while the short DRW timescale is already unresolved. 
%The current time resolution prevents us from statistically distinguishing these components as they reach the bin size.
 
Local accretion-disk simulations show that the effective viscosity depends sensitively on the net vertical magnetic flux \citep{2016MNRAS.460.3488S, 2025OJAp....8E..39S, 2025arXiv250512671G}. In the presence of a net vertical flux, strong magnetic pressure support produces $\alpha \sim 0.1$–$1$ in local simulations \citep[e.g.,][]{1995ApJ...440..742H, 2016MNRAS.457..857S, 2025arXiv250512671G}. General relativistic MHD simulations of thick, strongly magnetized accretion flows with diverse magnetic topologies also find large values of $\alpha$. These include magnetically arrested disks (MAD; \citealt{2003PASJ...55L..69N, 2012ApJ...757L..20N}), disks threaded by net vertical flux \citep{2016MNRAS.460.3488S}, and toroidally magnetized flows \citep{2025arXiv250512671G}. Such simulations typically obtain $\alpha \sim 0.1$–$0.3$, reaching $\sim 1$ in extreme cases \citep{2013ApJ...767...30B, 2013MNRAS.428.2255P, 2025OJAp....8E..39S, 2025arXiv250512671G}. By contrast, simulations without net vertical flux consistently yield $\alpha \sim 0.01$ \citep{1996ApJ...463..656S, 2010ApJ...713...52D, 2011ApJ...730...94S}, comparable to the canonical value assumed in the classical $\alpha$-disk model \citep{1973A&A....24..337S}.

%Independent observational constraints from dwarf novae and X-ray transient outbursts favor a similar viscosity range, $\alpha \sim 0.1$--$0.3$ \citep{2007MNRAS.376.1740K, 2018Natur.554...69T}. 
% \textcolor{blue}{Under the adopted physical interpretation, the ratio of the stochastic damping timescales is consistent with an effective viscosity parameter of order $\alpha\sim0.1$ in Swift J1727.8$-$1613.}
% Such a high-viscosity state suggests that the accretion flow in Swift J1727.8-1613 may be dominated by strong magnetic stresses and may likely be in a magnetically arrested disk (MAD) state \citep{2003PASJ...55L..69N, 2012ApJ...757L..20N}. 
% The convergence of the damping timescales toward the 0.1~s sampling limit near the jet ejection point (MJD 60206) further implies that the disk enters an extremely magneto-active state.
Under the adopted interpretation, the ratio of the stochastic damping timescales in Swift J1727.8-1613 is consistent with an effective viscosity parameter of order $\alpha\sim0.1$. This value is compatible with strongly magnetized accretion-flow models \citep{2003PASJ...55L..69N, 2012ApJ...757L..20N}.

\subsection{Intermittency and spatial distribution of local turbulence}

The observation that the short-timescale DRW component is negligible in a large fraction of GTIs for the LE and ME bands during the hard state, while persistently required in the HE band, provides a potential diagnostic for the spatial distribution and intermittency of magneto-hydrodynamic turbulence. In our proposed scenario, this short-timescale component traces local, high-frequency turbulent cascades and rapid magnetic reconnection.

During the hard state, the LE and ME emission may contain a larger contribution from relatively cooler and more extended regions of the disk--corona system. The high proportion of LE and ME GTIs lacking the short-timescale DRW implies that small-scale, high-frequency turbulence is significantly suppressed or highly intermittent in these outer regions. This could occur if these regions are dominated by a highly ordered magnetic field topology, where strong magnetic tension largely prevents the continuous development of rapid turbulent cascades, allowing them to occur only sporadically.

Conversely, the HE emission primarily originates from the innermost, highly Comptonized region of the corona. The persistent requirement of the short-timescale DRW in almost all HE GTIs suggests that this innermost zone maintains continuous, active magnetic reconnection and localized turbulence, even when the outer regions appear more magnetically rigid. This energy-dependent fractional behavior hints at a tomographic view of the magnetic activity in the accretion flow, which warrants further detailed investigation.

\section{Conclusion}
\label{s:5}
We have presented a novel framework for characterizing the stochastic and oscillatory dynamics of the black hole X-ray binary Swift J1727.8-1613. By employing GP inference with a physically motivated composite kernel (SHO + 2DRW), we successfully decomposed the multi-band light curves from \textit{Insight}-HXMT into distinct oscillatory and stochastic variability components. Our results provide a link between observational timing features and the underlying magneto-hydrodynamics of the accretion flow. The main conclusions are summarized as follows:  
\begin{itemize}     
 \item \textbf{Evolution of the Magneto-Active Cavity:} The GP modeling reveals a systematic evolution of the oscillator parameters during the outburst. The increase in $Q$ from $ \sim 3$ to $ \sim 10$, coupled with the rising QPO frequencies, suggests a transition toward a more rigid and coherent magnetic structure. We interpret this as the signature of a magnetically confined disk-coronal cavity that contracts as the system approaches the state transition.   
 
 \item  \textbf{Viscosity Constraints in Magnetized Flows:} The short DRW damping timescale is anti-correlated with the QPO frequency. This trend is consistent with our proposed stochastic magneto-active dynamics scenario. Under the proposed mapping of the DRW timescales to the dynamical and thermal timescales of the accretion flow, their ratio provides a physically motivated proxy for the effective viscosity parameter $\alpha$. The observed ratio is of order 0.1 and is therefore consistent with $\alpha\sim0.1$ within this interpretation.
 
\item  \textbf{Evolution of Characteristic Timescales:} We observed a significant decrease in both $\tau_{\mathrm{long}}^{\mathrm{DRW}}$ and $\tau_{\mathrm{short}}^{\mathrm{DRW}}$ as the system approached MJD 60206. Near the onset of the relativistic jet, the long DRW timescale decreases toward the 0.1 s sampling limit, while the short DRW timescale is already unresolved. The rapid decrease of the resolved characteristic timescales indicates a substantial change in the stochastic variability of the accretion flow around the jet-ejection epoch. The current time resolution prevents us from confirming the simultaneous alignment of these dynamical scales with the jet launching event.
\end{itemize}

Overall, this work demonstrates that GP inference is a powerful tool for bridging the gap between X-ray timing data and fundamental accretion physics. Future high-cadence observations across broader energy bands \citep[e.g., eXTP,][]{2025SCPMA..6819502Z, 2025SCPMA..6819506Y} will be instrumental in testing the universality of these magneto-dynamic scaling relations in other black hole candidates.

\begin{acknowledgements}
%%%%%%%%%%%%%%%%%%%%%%%%%
D.Y. acknowledges support from the National Natural
Science Foundation of China (grant No. 12393852) and
the Yunnan Provincial Science and Technology Department
Foundation (grant No. 202601AT070175). The observational data used in this study were obtained from the Insight-HXMT satellite, which is a mission jointly funded by the China National Space Administration (CNSA) and the Chinese Academy of Sciences (CAS).
\end{acknowledgements}

\bibliographystyle{aa} % use aa.bst
\bibliography{MyBiblio} % References in MyBiblio.bib; run BibTeX after LaTeX

@ARTICLE{2024MNRAS.529.4624Y,
       author = {{Yu}, Wei and {Bu}, Qing-Cui and {Zhang}, Shuang-Nan and {Liu}, He-Xin and {Zhang}, Liang and {Ducci}, Lorenzo and {Tao}, Lian and {Santangelo}, Andrea and {Doroshenko}, Victor and {Huang}, Yue and {Yang}, Zi-Xu and {Qu}, Jin-Lu},
        title = "{Timing analysis of the newly discovered black hole candidate Swift J1727.8-1613 with Insight-HXMT}",
      journal = {\mnras},
         year = 2024,
        month = apr,
       volume = {529},
       number = {4},
        pages = {4624-4632},
          doi = {10.1093/mnras/stae835},
archivePrefix = {arXiv},
       eprint = {2403.13127},
 primaryClass = {astro-ph.HE},
       adsurl = {https://ui.adsabs.harvard.edu/abs/2024MNRAS.529.4624Y}
}

@ARTICLE{2024ApJ...966L..35S,
       author = {{Svoboda}, Ji{\v{r}}{\'\i} and {Dov{\v{c}}iak}, Michal and {Steiner}, James F. and {Kaaret}, Philip and {Podgorn{\'y}}, Jakub and {Poutanen}, Juri and {Veledina}, Alexandra and {Muleri}, Fabio and {Taverna}, Roberto and {Krawczynski}, Henric and {Brigitte}, Ma{\"\i}mouna and {Datta}, Sudeb Ranjan and {Bianchi}, Stefano and {Mu{\~n}oz-Darias}, Teo and {Negro}, Michela and {Rodriguez Cavero}, Nicole and {Castro Segura}, Noel and {Bollemeijer}, Niek and {Garc{\'\i}a}, Javier A. and {Ingram}, Adam and {Matt}, Giorgio and {Nathan}, Edward and {Weisskopf}, Martin C. and {Altamirano}, Diego and {Baldini}, Luca and {Capitanio}, Fiamma and {Egron}, Elise and {Emami}, Razieh and {Hu}, Kun and {Marra}, Lorenzo and {Mastroserio}, Guglielmo and {Petrucci}, Pierre-Olivier and {Ratheesh}, Ajay and {Soffitta}, Paolo and {Tombesi}, Francesco and {Yang}, Yi-Jung and {Zhang}, Yuexin},
        title = "{Dramatic Drop in the X-Ray Polarization of Swift J1727.8{\textendash}1613 in the Soft Spectral State}",
      journal = {\apjl},
         year = 2024,
        month = may,
       volume = {966},
       number = {2},
          eid = {L35},
        pages = {L35},
          doi = {10.3847/2041-8213/ad402e},
archivePrefix = {arXiv},
       eprint = {2403.04689},
 primaryClass = {astro-ph.HE},
       adsurl = {https://ui.adsabs.harvard.edu/abs/2024ApJ...966L..35S}
}

@ARTICLE{2023GCN.33465....1B,
       author = {{Burrows}, D.~N. and {Gropp}, J.~D. and {Osborne}, J.~P. and {Page}, K.~L. and {D'Elia}, V. and {Sbarufatti}, B. and {D'Ai}, A. and {Dichiara}, S. and {Evans}, P.~A. and {Swift-XRT Team}},
        title = "{GRB 230307A: Swift-XRT afterglow detection}",
      journal = {GRB Coordinates Network},
         year = 2023,
        month = mar,
       volume = {33465},
        pages = {1},
       adsurl = {https://ui.adsabs.harvard.edu/abs/2023GCN.33465....1B}
}

@ARTICLE{2025A&A...693A.129M,
       author = {{Mata S{\'a}nchez}, D. and {Torres}, M.~A.~P. and {Casares}, J. and {Mu{\~n}oz-Darias}, T. and {Armas Padilla}, M. and {Yanes-Rizo}, I.~V.},
        title = "{Dynamical confirmation of a black hole in the X-ray transient Swift J1727.8‑1613}",
      journal = {\aap},
         year = 2025,
        month = jan,
       volume = {693},
          eid = {A129},
        pages = {A129},
          doi = {10.1051/0004-6361/202451960},
archivePrefix = {arXiv},
       eprint = {2408.13310},
 primaryClass = {astro-ph.HE},
       adsurl = {https://ui.adsabs.harvard.edu/abs/2025A&A...693A.129M}
}

@ARTICLE{2025ApJ...994..243B,
       author = {{Burridge}, Benjamin J. and {Miller-Jones}, James C.~A. and {Bahramian}, Arash and {Prabu}, Steve R. and {Streeter}, Reagan and {Castro Segura}, Noel and {Corral-Santana}, Jes{\'u}s M. and {Knigge}, Christian and {Zdziarski}, Andrzej and {Mata S{\'a}nchez}, Daniel and {Tremou}, Evangelia and {Carotenuto}, Francesco and {Fender}, Rob and {Saikia}, Payaswini},
        title = "{On the Distance to the Black Hole X-Ray Binary Swift J1727.8─1613}",
      journal = {\apj},
         year = 2025,
        month = dec,
       volume = {994},
       number = {2},
          eid = {243},
        pages = {243},
          doi = {10.3847/1538-4357/ae11a0},
archivePrefix = {arXiv},
       eprint = {2502.06448},
 primaryClass = {astro-ph.HE},
       adsurl = {https://ui.adsabs.harvard.edu/abs/2025ApJ...994..243B}
}

@ARTICLE{2025ApJ...983...23C,
       author = {{Cao}, Jia-Ying and {Liao}, Jin-Yuan and {Zhang}, Shuang-Nan and {Feng}, Hua and {Qu}, Jin-Lu and {Zhang}, Liang and {Liu}, He-Xin and {Yu}, Wei and {Zhao}, Qing-Chang and {Peng}, Jing-Qiang and {Ge}, Ming-Yu and {Tao}, Lian and {Xu}, Yan-Jun and {Zhang}, Shu and {Yang}, Zi-Xu},
        title = "{Spectral Analysis of the X-Ray Flares in the 2023 Outburst of the New Black Binary Transient Swift J1727.8{\textendash}1613 Observed with Insight-HXMT}",
      journal = {\apj},
         year = 2025,
        month = apr,
       volume = {983},
       number = {1},
          eid = {23},
        pages = {23},
          doi = {10.3847/1538-4357/adbd0f},
archivePrefix = {arXiv},
       eprint = {2503.05411},
 primaryClass = {astro-ph.HE},
       adsurl = {https://ui.adsabs.harvard.edu/abs/2025ApJ...983...23C}
}

@ARTICLE{2025ApJ...986....3L,
       author = {{Liao}, Jie and {Chang}, Ning and {Cui}, Lang and {Jiang}, Pengfei and {Mou}, Didong and {Huang}, Yongfeng and {An}, Tao and {Ho}, Luis C. and {Feng}, Hua and {Fu}, Yu-Cong and {Cao}, Hongmin and {Tripathi}, Ashutosh and {Liu}, Xiang},
        title = "{Tracking the Jet-like Corona of Black Hole Swift J1727.8‑1613 during a Flare State through Type-C Quasiperiodic Oscillations}",
      journal = {\apj},
         year = 2025,
        month = jun,
       volume = {986},
       number = {1},
          eid = {3},
        pages = {3},
          doi = {10.3847/1538-4357/add264},
archivePrefix = {arXiv},
       eprint = {2410.06574},
 primaryClass = {astro-ph.HE},
       adsurl = {https://ui.adsabs.harvard.edu/abs/2025ApJ...986....3L}
}

@ARTICLE{2023ApJ...958L..16V,
       author = {{Veledina}, Alexandra and {Muleri}, Fabio and {Dov{\v{c}}iak}, Michal and {Poutanen}, Juri and {Ratheesh}, Ajay and {Capitanio}, Fiamma and {Matt}, Giorgio and {Soffitta}, Paolo and {Tennant}, Allyn F. and {Negro}, Michela and {Kaaret}, Philip and {Costa}, Enrico and {Ingram}, Adam and {Svoboda}, Ji{\v{r}}{\'\i} and {Krawczynski}, Henric and {Bianchi}, Stefano and {Steiner}, James F. and {Garc{\'\i}a}, Javier A. and {Kravtsov}, Vadim and {Nitindala}, Anagha P. and {Ewing}, Melissa and {Mastroserio}, Guglielmo and {Marinucci}, Andrea and {Ursini}, Francesco and {Tombesi}, Francesco and {Tsygankov}, Sergey S. and {Yang}, Yi-Jung and {Weisskopf}, Martin C. and {Trushkin}, Sergei A. and {Egron}, Elise and {Iacolina}, Maria Noemi and {Pilia}, Maura and {Marra}, Lorenzo and {Miku{\v{s}}incov{\'a}}, Romana and {Nathan}, Edward and {Parra}, Maxime and {Petrucci}, Pierre-Olivier and {Podgorn{\'y}}, Jakub and {Tugliani}, Stefano and {Zane}, Silvia and {Zhang}, Wenda and {Agudo}, Iv{\'a}n and {Antonelli}, Lucio A. and {Bachetti}, Matteo and {Baldini}, Luca and {Baumgartner}, Wayne H. and {Bellazzini}, Ronaldo and {Bongiorno}, Stephen D. and {Bonino}, Raffaella and {Brez}, Alessandro and {Bucciantini}, Niccol{\`o} and {Castellano}, Simone and {Cavazzuti}, Elisabetta and {Chen}, Chien-Ting and {Ciprini}, Stefano and {De Rosa}, Alessandra and {Del Monte}, Ettore and {Di Gesu}, Laura and {Di Lalla}, Niccol{\`o} and {Di Marco}, Alessandro and {Donnarumma}, Immacolata and {Doroshenko}, Victor and {Ehlert}, Steven R. and {Enoto}, Teruaki and {Evangelista}, Yuri and {Fabiani}, Sergio and {Ferrazzoli}, Riccardo and {Gunji}, Shuichi and {Hayashida}, Kiyoshi and {Heyl}, Jeremy and {Iwakiri}, Wataru and {Jorstad}, Svetlana G. and {Karas}, Vladimir and {Kislat}, Fabian and {Kitaguchi}, Takao and {Kolodziejczak}, Jeffery J. and {La Monaca}, Fabio and {Latronico}, Luca and {Liodakis}, Ioannis and {Maldera}, Simone and {Manfreda}, Alberto and {Marin}, Fr{\'e}d{\'e}ric and {Marscher}, Alan P. and {Marshall}, Herman L. and {Massaro}, Francesco and {Mitsuishi}, Ikuyuki and {Mizuno}, Tsunefumi and {Ng}, Chi-Yung and {O'Dell}, Stephen L. and {Omodei}, Nicola and {Oppedisano}, Chiara and {Papitto}, Alessandro and {Pavlov}, George G. and {Peirson}, Abel L. and {Perri}, Matteo and {Pesce-Rollins}, Melissa and {Possenti}, Andrea and {Puccetti}, Simonetta and {Ramsey}, Brian D. and {Rankin}, John and {Roberts}, Oliver J. and {Romani}, Roger W. and {Sgr{\`o}}, Carmelo and {Slane}, Patrick and {Spandre}, Gloria and {Swartz}, Douglas A. and {Tamagawa}, Toru and {Tavecchio}, Fabrizio and {Taverna}, Roberto and {Tawara}, Yuzuru and {Thomas}, Nicholas E. and {Trois}, Alessio and {Turolla}, Roberto and {Vink}, Jacco and {Wu}, Kinwah and {Xie}, Fei},
        title = "{Discovery of X-Ray Polarization from the Black Hole Transient Swift J1727.8-1613}",
      journal = {\apjl},
         year = 2023,
        month = nov,
       volume = {958},
       number = {1},
          eid = {L16},
        pages = {L16},
          doi = {10.3847/2041-8213/ad0781},
archivePrefix = {arXiv},
       eprint = {2309.15928},
 primaryClass = {astro-ph.HE},
       adsurl = {https://ui.adsabs.harvard.edu/abs/2023ApJ...958L..16V}
}

@ARTICLE{2024A&A...686L..12P,
       author = {{Podgorn{\'y}}, J. and {Svoboda}, J. and {Dov{\v{c}}iak}, M. and {Veledina}, A. and {Poutanen}, J. and {Kaaret}, P. and {Bianchi}, S. and {Ingram}, A. and {Capitanio}, F. and {Datta}, S.~R. and {Egron}, E. and {Krawczynski}, H. and {Matt}, G. and {Muleri}, F. and {Petrucci}, P.-O. and {Russell}, T.~D. and {Steiner}, J.~F. and {Bollemeijer}, N. and {Brigitte}, M. and {Castro Segura}, N. and {Emami}, R. and {Garc{\'\i}a}, J.~A. and {Hu}, K. and {Iacolina}, M.~N. and {Kravtsov}, V. and {Marra}, L. and {Mastroserio}, G. and {Mu{\~n}oz-Darias}, T. and {Nathan}, E. and {Negro}, M. and {Ratheesh}, A. and {Rodriguez Cavero}, N. and {Taverna}, R. and {Tombesi}, F. and {Yang}, Y.~J. and {Zhang}, W. and {Zhang}, Y.},
        title = "{Recovery of the X-ray polarisation of Swift J1727.8‒1613 after the soft-to-hard spectral transition}",
      journal = {\aap},
         year = 2024,
        month = jun,
       volume = {686},
          eid = {L12},
        pages = {L12},
          doi = {10.1051/0004-6361/202450566},
archivePrefix = {arXiv},
       eprint = {2404.19601},
 primaryClass = {astro-ph.HE},
       adsurl = {https://ui.adsabs.harvard.edu/abs/2024A&A...686L..12P}
}

@ARTICLE{2024ApJ...971L...9W,
       author = {{Wood}, Callan M. and {Miller-Jones}, James C.~A. and {Bahramian}, Arash and {Tingay}, Steven J. and {Prabu}, Steve and {Russell}, Thomas D. and {Atri}, Pikky and {Carotenuto}, Francesco and {Altamirano}, Diego and {Motta}, Sara E. and {Hyland}, Lucas and {Reynolds}, Cormac and {Weston}, Stuart and {Fender}, Rob and {K{\"o}rding}, Elmar and {Maitra}, Dipankar and {Markoff}, Sera and {Migliari}, Simone and {Russell}, David M. and {Sarazin}, Craig L. and {Sivakoff}, Gregory R. and {Soria}, Roberto and {Tetarenko}, Alexandra J. and {Tudose}, Valeriu},
        title = "{Swift J1727.8{\textendash}1613 Has the Largest Resolved Continuous Jet Ever Seen in an X-Ray Binary}",
      journal = {\apjl},
         year = 2024,
        month = aug,
       volume = {971},
       number = {1},
          eid = {L9},
        pages = {L9},
          doi = {10.3847/2041-8213/ad6572},
archivePrefix = {arXiv},
       eprint = {2405.12370},
 primaryClass = {astro-ph.HE},
       adsurl = {https://ui.adsabs.harvard.edu/abs/2024ApJ...971L...9W}
}

@ARTICLE{2025ApJ...984L..53W,
       author = {{Wood}, Callan M. and {Miller-Jones}, James C.~A. and {Bahramian}, Arash and {Tingay}, Steven J. and {Liu}, He-Xin and {Altamirano}, Diego and {Fender}, Rob and {K{\"o}rding}, Elmar and {Maitra}, Dipankar and {Markoff}, Sera and {Russell}, David M. and {Russell}, Thomas D. and {Sarazin}, Craig L. and {Sivakoff}, Gregory R. and {Soria}, Roberto and {Tetarenko}, Alexandra J. and {Tudose}, Valeriu},
        title = "{The Ejection of Transient Jets in Swift J1727.8‑1613 Revealed by Time-dependent Visibility Modeling}",
      journal = {\apjl},
         year = 2025,
        month = may,
       volume = {984},
       number = {2},
          eid = {L53},
        pages = {L53},
          doi = {10.3847/2041-8213/adc9b3},
archivePrefix = {arXiv},
       eprint = {2503.03073},
 primaryClass = {astro-ph.HE},
       adsurl = {https://ui.adsabs.harvard.edu/abs/2025ApJ...984L..53W}
}

@ARTICLE{2025A&A...703A.134W,
       author = {{Wang}, Yiran and {Ma}, Ruican and {Zhang}, Haiyun and {Yan}, Dahai},
        title = "{Gaussian process analysis of type-B quasiperiodic oscillations in the black hole X-ray binary MAXI J1348─630}",
      journal = {\aap},
         year = 2025,
        month = nov,
       volume = {703},
          eid = {A134},
        pages = {A134},
          doi = {10.1051/0004-6361/202555561},
archivePrefix = {arXiv},
       eprint = {2510.05587},
 primaryClass = {astro-ph.HE},
       adsurl = {https://ui.adsabs.harvard.edu/abs/2025A&A...703A.134W}
}

@ARTICLE{2020SCPMA..6349502Z,
       author = {{Zhang}, Shuang-Nan and {Li}, TiPei and {Lu}, FangJun and {Song}, LiMing and {Xu}, YuPeng and {Liu}, CongZhan and {Chen}, Yong and {Cao}, XueLei and {Bu}, QingCui and {Chang}, Zhi and {Chen}, Gang and {Chen}, Li and {Chen}, TianXiang and {Chen}, YiBao and {Chen}, YuPeng and {Cui}, Wei and {Cui}, WeiWei and {Deng}, JingKang and {Dong}, YongWei and {Du}, YuanYuan and {Fu}, MinXue and {Gao}, GuanHua and {Gao}, He and {Gao}, Min and {Ge}, MingYu and {Gu}, YuDong and {Guan}, Ju and {Gungor}, Can and {Guo}, ChengCheng and {Han}, DaWei and {Hu}, Wei and {Huang}, Yue and {Huo}, Jia and {Jia}, ShuMei and {Jiang}, LuHua and {Jiang}, WeiChun and {Jin}, Jing and {Jin}, YongJie and {Li}, Bing and {Li}, ChengKui and {Li}, Gang and {Li}, MaoShun and {Li}, Wei and {Li}, Xian and {Li}, XiaoBo and {Li}, XuFang and {Li}, YanGuo and {Li}, ZiJian and {Li}, ZhengWei and {Liang}, XiaoHua and {Liao}, JinYuan and {Liu}, GuoQing and {Liu}, HongWei and {Liu}, ShaoZhen and {Liu}, XiaoJing and {Liu}, Yuan and {Liu}, YiNong and {Lu}, Bo and {Lu}, XueFeng and {Luo}, Tao and {Ma}, Xiang and {Meng}, Bin and {Nang}, Yi and {Nie}, JianYin and {Ou}, Ge and {Qu}, JinLu and {Sai}, Na and {Shang}, RenCheng and {Shen}, GuoHong and {Sun}, Liang and {Tan}, Ying and {Tao}, Lian and {Tuo}, YouLi and {Wang}, Chen and {Wang}, ChunQin and {Wang}, GuoFeng and {Wang}, HuanYu and {Wang}, Juan and {Wang}, WenShuai and {Wang}, YuSa and {Wen}, XiangYang and {Wu}, BaiYang and {Wu}, BoBing and {Wu}, Mei and {Xiao}, GuangCheng and {Xiong}, ShaoLin and {Yan}, LinLi and {Yang}, JiaWei and {Yang}, Sheng and {Yang}, YanJi and {Yi}, QiBin and {Yuan}, Bin and {Zhang}, AiMei and {Zhang}, ChunLei and {Zhang}, ChengMo and {Zhang}, Fan and {Zhang}, HongMei and {Zhang}, Juan and {Zhang}, Qiang and {Zhang}, ShenYi and {Zhang}, Shu and {Zhang}, Tong and {Zhang}, WanChang and {Zhang}, Wei and {Zhang}, WenZhao and {Zhang}, Yi and {Zhang}, YiFei and {Zhang}, YongJie and {Zhang}, Yue and {Zhang}, Zhao and {Zhang}, Zhi and {Zhang}, ZiLiang and {Zhao}, HaiSheng and {Zhao}, XiaoFan and {Zheng}, ShiJie and {Zhou}, JianFeng and {Zhu}, YuXuan and {Zhu}, Yue and {Zhuang}, RenLin and {The Insight-HXMT team}},
        title = "{Overview to the Hard X-ray Modulation Telescope (Insight-HXMT) Satellite}",
      journal = {Science China Physics, Mechanics, and Astronomy},
         year = 2020,
        month = apr,
       volume = {63},
       number = {4},
          eid = {249502},
        pages = {249502},
          doi = {10.1007/s11433-019-1432-6},
archivePrefix = {arXiv},
       eprint = {1910.09613},
 primaryClass = {astro-ph.IM},
       adsurl = {https://ui.adsabs.harvard.edu/abs/2020SCPMA..6349502Z}
}

@ARTICLE{2020SCPMA..6349505C,
       author = {{Chen}, Yong and {Cui}, WeiWei and {Li}, Wei and {Wang}, Juan and {Xu}, YuPeng and {Lu}, FangJun and {Wang}, YuSa and {Chen}, TianXiang and {Han}, DaWei and {Hu}, Wei and {Zhang}, Yi and {Huo}, Jia and {Yang}, YanJi and {Li}, MaoShun and {Lu}, Bo and {Zhang}, ZiLiang and {Li}, TiPei and {Zhang}, ShuangNan and {Xiong}, ShaoLin and {Zhang}, Shu and {Xue}, RongFeng and {Zhao}, XiaoFan and {Zhu}, Yue and {Zhu}, YuXuan and {Liu}, HongWei and {Yang}, YiJung and {Zhang}, Fan},
        title = "{The Low Energy X-ray telescope (LE) onboard the Insight-HXMT astronomy satellite}",
      journal = {Science China Physics, Mechanics, and Astronomy},
         year = 2020,
        month = apr,
       volume = {63},
       number = {4},
          eid = {249505},
        pages = {249505},
          doi = {10.1007/s11433-019-1469-5},
archivePrefix = {arXiv},
       eprint = {1910.08319},
 primaryClass = {astro-ph.IM},
       adsurl = {https://ui.adsabs.harvard.edu/abs/2020SCPMA..6349505C}
}

@ARTICLE{2020SCPMA..6349504C,
       author = {{Cao}, XueLei and {Jiang}, WeiChun and {Meng}, Bin and {Zhang}, WanChang and {Luo}, Tao and {Yang}, Sheng and {Zhang}, ChunLei and {Gu}, YuDong and {Sun}, Liang and {Liu}, XiaoJing and {Yang}, JiaWei and {Li}, Xian and {Tan}, Ying and {Liu}, ShaoZhen and {Du}, YuanYuan and {Lu}, FangJun and {Xu}, YuPeng and {Guan}, Ju and {Zhang}, ShuangNan and {Wang}, HuanYu and {Li}, TiPei and {Zhang}, ChengMo and {Wen}, XiangYang and {Qu}, JinLu and {Song}, LiMing and {Li}, XiaoBo and {Ge}, MingYu and {Zhou}, YuPeng and {Xiong}, ShaoLin and {Zhang}, Shu and {Zhang}, YongJie and {Cheng}, ZeHao and {Zhang}, Fei and {Li}, MaoShun and {Liang}, XiaoHua and {Gao}, Min and {Yang}, EnBo and {Liu}, XiaoHang and {Liu}, HongWei and {Yang}, YiJung and {Zhang}, Fan},
        title = "{The Medium Energy X-ray telescope (ME) onboard the Insight-HXMT astronomy satellite}",
      journal = {Science China Physics, Mechanics, and Astronomy},
         year = 2020,
        month = apr,
       volume = {63},
       number = {4},
          eid = {249504},
        pages = {249504},
          doi = {10.1007/s11433-019-1506-1},
archivePrefix = {arXiv},
       eprint = {1910.04451},
 primaryClass = {astro-ph.IM},
       adsurl = {https://ui.adsabs.harvard.edu/abs/2020SCPMA..6349504C}
}

@ARTICLE{2020SCPMA..6349503L,
       author = {{Liu}, CongZhan and {Zhang}, YiFei and {Li}, XuFang and {Lu}, XueFeng and {Chang}, Zhi and {Li}, ZhengWei and {Zhang}, AiMei and {Jin}, YongJie and {Yu}, HuiMing and {Zhang}, Zhao and {Fu}, MinXue and {Chen}, YiBao and {Ji}, JianFeng and {Xu}, YuPeng and {Deng}, JingKang and {Shang}, RenCheng and {Liu}, GuoQing and {Lu}, FangJun and {Zhang}, ShuangNan and {Dong}, YongWei and {Li}, TiPei and {Wu}, Mei and {Li}, YanGuo and {Wang}, HuanYu and {Wu}, BoBing and {Zhang}, YongJie and {Zhang}, Zhi and {Xiong}, ShaoLin and {Liu}, Yuan and {Zhang}, Shu and {Liu}, HongWei and {Yang}, YiJung and {Zhang}, Fan},
        title = "{The High Energy X-ray telescope (HE) onboard the Insight-HXMT astronomy satellite}",
      journal = {Science China Physics, Mechanics, and Astronomy},
         year = 2020,
        month = apr,
       volume = {63},
       number = {4},
          eid = {249503},
        pages = {249503},
          doi = {10.1007/s11433-019-1486-x},
archivePrefix = {arXiv},
       eprint = {1910.04955},
 primaryClass = {astro-ph.IM},
       adsurl = {https://ui.adsabs.harvard.edu/abs/2020SCPMA..6349503L}
}

@ARTICLE{2025ApJ...993...40X,
       author = {{Xu}, Sai-En and {You}, Bei and {Long}, Yi and {He}, Han},
        title = "{Temporal Evolution of Quasiperiodic Oscillations in an Accreting Black Hole Swift J1727.8-1613: Coevolution of the Disk─Corona during the State Transition}",
      journal = {\apj},
         year = 2025,
        month = nov,
       volume = {993},
       number = {1},
          eid = {40},
        pages = {40},
          doi = {10.3847/1538-4357/ae058a},
archivePrefix = {arXiv},
       eprint = {2506.21131},
 primaryClass = {astro-ph.HE},
       adsurl = {https://ui.adsabs.harvard.edu/abs/2025ApJ...993...40X}
}

@ARTICLE{2024ApJ...968...76I,
       author = {{Ingram}, Adam and {Bollemeijer}, Niek and {Veledina}, Alexandra and {Dov{\v{c}}iak}, Michal and {Poutanen}, Juri and {Egron}, Elise and {Russell}, Thomas D. and {Trushkin}, Sergei A. and {Negro}, Michela and {Ratheesh}, Ajay and {Capitanio}, Fiamma and {Connors}, Riley and {Neilsen}, Joseph and {Kraus}, Alexander and {Iacolina}, Maria Noemi and {Pellizzoni}, Alberto and {Pilia}, Maura and {Carotenuto}, Francesco and {Matt}, Giorgio and {Mastroserio}, Guglielmo and {Kaaret}, Philip and {Bianchi}, Stefano and {Garc{\'\i}a}, Javier A. and {Bachetti}, Matteo and {Wu}, Kinwah and {Costa}, Enrico and {Ewing}, Melissa and {Kravtsov}, Vadim and {Krawczynski}, Henric and {Loktev}, Vladislav and {Marinucci}, Andrea and {Marra}, Lorenzo and {Miku{\v{s}}incov{\'a}}, Romana and {Nathan}, Edward and {Parra}, Maxime and {Petrucci}, Pierre-Olivier and {Righini}, Simona and {Soffitta}, Paolo and {Steiner}, James F. and {Svoboda}, Ji{\v{r}}{\'\i} and {Tombesi}, Francesco and {Tugliani}, Stefano and {Ursini}, Francesco and {Yang}, Yi-Jung and {Zane}, Silvia and {Zhang}, Wenda and {Agudo}, Iv{\'a}n and {Antonelli}, Lucio A. and {Baldini}, Luca and {Baumgartner}, Wayne H. and {Bellazzini}, Ronaldo and {Bongiorno}, Stephen D. and {Bonino}, Raffaella and {Brez}, Alessandro and {Bucciantini}, Niccol{\`o} and {Castellano}, Simone and {Cavazzuti}, Elisabetta and {Chen}, Chien-Ting and {Ciprini}, Stefano and {De Rosa}, Alessandra and {Del Monte}, Ettore and {Di Gesu}, Laura and {Di Lalla}, Niccol{\`o} and {Di Marco}, Alessandro and {Donnarumma}, Immacolata and {Doroshenko}, Victor and {Ehlert}, Steven R. and {Enoto}, Teruaki and {Evangelista}, Yuri and {Fabiani}, Sergio and {Ferrazzoli}, Riccardo and {Gunji}, Shuichi and {Hayashida}, Kiyoshi and {Heyl}, Jeremy and {Iwakiri}, Wataru and {Jorstad}, Svetlana G. and {Karas}, Vladimir and {Kislat}, Fabian and {Kitaguchi}, Takao and {Kolodziejczak}, Jeffery J. and {La Monaca}, Fabio and {Latronico}, Luca and {Liodakis}, Ioannis and {Maldera}, Simone and {Manfreda}, Alberto and {Marin}, Fr{\'e}d{\'e}ric and {Marscher}, Alan P. and {Marshall}, Herman L. and {Massaro}, Francesco and {Mitsuishi}, Ikuyuki and {Mizuno}, Tsunefumi and {Muleri}, Fabio and {Ng}, Chi-Yung and {O'Dell}, Stephen L. and {Omodei}, Nicola and {Oppedisano}, Chiara and {Papitto}, Alessandro and {Pavlov}, George G. and {Peirson}, Abel L. and {Perri}, Matteo and {Pesce-Rollins}, Melissa and {Possenti}, Andrea and {Puccetti}, Simonetta and {Ramsey}, Brian D. and {Rankin}, John and {Roberts}, Oliver J. and {Romani}, Roger W. and {Sgr{\`o}}, Carmelo and {Slane}, Patrick and {Spandre}, Gloria and {Swartz}, Douglas A. and {Tamagawa}, Toru and {Tavecchio}, Fabrizio and {Taverna}, Roberto and {Tawara}, Yuzuru and {Tennant}, Allyn F. and {Thomas}, Nicholas E. and {Trois}, Alessio and {Tsygankov}, Sergey S. and {Turolla}, Roberto and {Vink}, Jacco and {Weisskopf}, Martin C. and {Xie}, Fei and {IXPE Collaboration}},
        title = "{Tracking the X-Ray Polarization of the Black Hole Transient Swift J1727.8{\textendash}1613 during a State Transition}",
      journal = {\apj},
         year = 2024,
        month = jun,
       volume = {968},
       number = {2},
          eid = {76},
        pages = {76},
          doi = {10.3847/1538-4357/ad3faf},
archivePrefix = {arXiv},
       eprint = {2311.05497},
 primaryClass = {astro-ph.HE},
       adsurl = {https://ui.adsabs.harvard.edu/abs/2024ApJ...968...76I}
}

@ARTICLE{2024ATel16541....1P,
       author = {{Podgorny}, Jakub and {Svoboda}, Jiri and {Dovciak}, Michal},
        title = "{MAXI observations of Swift J1727.8-1613 indicate transition of the source to the low/hard state}",
      journal = {The Astronomer's Telegram},
         year = 2024,
        month = mar,
       volume = {16541},
        pages = {1},
       adsurl = {https://ui.adsabs.harvard.edu/abs/2024ATel16541....1P}
}

@ARTICLE{2025MNRAS.542.1803H,
       author = {{Hughes}, A.~K. and {Carotenuto}, F. and {Russell}, T.~D. and {Tetarenko}, A.~J. and {Miller-Jones}, J.~C.~A. and {Plotkin}, R.~M. and {Bahramian}, A. and {Bright}, J.~S. and {Cowie}, F.~J. and {Crook-Mansour}, J. and {Fender}, R. and {Khaulsay}, J.~K. and {Kirby}, A. and {Jones}, S. and {McCollough}, M. and {Rao}, R. and {Sivakoff}, G.~R. and {Vrtilek}, S.~D. and {Williams-Baldwin}, D.~R.~A. and {Wood}, C.~M. and {Altamirano}, D. and {Casella}, P. and {Segura}, N. Castro and {Corbel}, S. and {Del Santo}, M. and {Echibur{\'u}-Trujillo}, C. and {van den Eijnden}, J. and {Gallo}, E. and {Gandhi}, P. and {Koljonen}, K.~I.~I. and {Maccarone}, T. and {Markoff}, S.~B. and {Motta}, S. and {Russell}, D.~M. and {Saikia}, P. and {Shaw}, A.~W. and {Soria}, R. and {Tomsick}, J.~A. and {Yu}, W. and {Zhang}, X.},
        title = "{The peculiar hard state behaviour of the black hole X-ray binary Swift J1727.8{\ensuremath{-}}1613}",
      journal = {\mnras},
         year = 2025,
        month = sep,
       volume = {542},
       number = {3},
        pages = {1803-1816},
          doi = {10.1093/mnras/staf1363},
archivePrefix = {arXiv},
       eprint = {2506.12387},
 primaryClass = {astro-ph.HE},
       adsurl = {https://ui.adsabs.harvard.edu/abs/2025MNRAS.542.1803H}
}

@ARTICLE{2025ApJ...988..109H,
       author = {{Hughes}, Andrew K. and {Carotenuto}, Francesco and {Russell}, Thomas D. and {Tetarenko}, Alexandra J. and {Miller-Jones}, James C.~A. and {Bahramian}, Arash and {Bright}, Joe S. and {Cowie}, Fraser J. and {Fender}, Rob and {Gurwell}, Mark A. and {Khaulsay}, Jasvinderjit K. and {Kirby}, Anastasia and {Jones}, Serena and {Lescure}, Elodie and {McCollough}, Michael and {Plotkin}, Richard M. and {Rao}, Ramprasad and {Vrtilek}, Saeqa D. and {Williams-Baldwin}, David R.~A. and {Wood}, Callan M. and {Sivakoff}, Gregory R. and {Altamirano}, Diego and {Casella}, Piergiorgio and {Corbel}, St{\'e}phane and {DeBoer}, David R. and {Del Santo}, Melania and {Echibur{\'u}-Trujillo}, Constanza and {Farah}, Wael and {Gandhi}, Poshak and {Koljonen}, Karri I.~I. and {Maccarone}, Thomas and {Matthews}, James H. and {Markoff}, Sera B. and {Pollak}, Alexander W. and {Russell}, David M. and {Saikia}, Payaswini and {Castro Segura}, Noel and {Shaw}, Aarran W. and {Siemion}, Andrew and {Soria}, Roberto and {Tomsick}, John A. and {van den Eijnden}, Jakob},
        title = "{Comprehensive Radio Monitoring of the Black Hole X-Ray Binary Swift J1727.8‑1613 during Its 2023{\textendash}2024 Outburst}",
      journal = {\apj},
         year = 2025,
        month = jul,
       volume = {988},
       number = {1},
          eid = {109},
        pages = {109},
          doi = {10.3847/1538-4357/ade2e6},
archivePrefix = {arXiv},
       eprint = {2506.07798},
 primaryClass = {astro-ph.HE},
       adsurl = {https://ui.adsabs.harvard.edu/abs/2025ApJ...988..109H}
}

@article{celerite2,
   author = {{Foreman-Mackey}, D.},
    title = "{Scalable Backpropagation for Gaussian Processes using Celerite}",
  journal = {Research Notes of the American Astronomical Society},
     year = 2018,
    month = feb,
   volume = 2,
   number = 1,
    pages = {31},
      doi = {10.3847/2515-5172/aaaf6c},
   adsurl = {http://adsabs.harvard.edu/abs/2018RNAAS...2a..31F}
}

@ARTICLE{2023ARA&A..61..329A,
       author = {{Aigrain}, Suzanne and {Foreman-Mackey}, Daniel},
        title = "{Gaussian Process Regression for Astronomical Time Series}",
      journal = {\araa},
         year = 2023,
        month = aug,
       volume = {61},
        pages = {329-371},
          doi = {10.1146/annurev-astro-052920-103508},
archivePrefix = {arXiv},
       eprint = {2209.08940},
 primaryClass = {astro-ph.IM},
       adsurl = {https://ui.adsabs.harvard.edu/abs/2023ARA&A..61..329A}
}

@ARTICLE{2017AJ....154..220F,
       author = {{Foreman-Mackey}, Daniel and {Agol}, Eric and {Ambikasaran}, Sivaram and {Angus}, Ruth},
        title = "{Fast and Scalable Gaussian Process Modeling with Applications to Astronomical Time Series}",
      journal = {\aj},
         year = 2017,
        month = dec,
       volume = {154},
       number = {6},
          eid = {220},
        pages = {220},
          doi = {10.3847/1538-3881/aa9332},
archivePrefix = {arXiv},
       eprint = {1703.09710},
 primaryClass = {astro-ph.IM},
       adsurl = {https://ui.adsabs.harvard.edu/abs/2017AJ....154..220F}
}

@ARTICLE{2022ApJ...930..157Z,
       author = {{Zhang}, Haiyun and {Yan}, Dahai and {Zhang}, Li},
        title = "{Characterizing the {\ensuremath{\gamma}}-Ray Variability of Active Galactic Nuclei with the Stochastic Process Method}",
      journal = {\apj},
         year = 2022,
        month = may,
       volume = {930},
       number = {2},
          eid = {157},
        pages = {157},
          doi = {10.3847/1538-4357/ac679e},
archivePrefix = {arXiv},
       eprint = {2204.09987},
 primaryClass = {astro-ph.HE},
       adsurl = {https://ui.adsabs.harvard.edu/abs/2022ApJ...930..157Z}
}

@ARTICLE{2023ApJ...944..103Z,
       author = {{Zhang}, Haiyun and {Yan}, Dahai and {Zhang}, Li},
        title = "{Gaussian Process Modeling Blazar Multiwavelength Variability: Indirectly Resolving Jet Structure}",
      journal = {\apj},
         year = 2023,
        month = feb,
       volume = {944},
       number = {1},
          eid = {103},
        pages = {103},
          doi = {10.3847/1538-4357/acafe5},
archivePrefix = {arXiv},
       eprint = {2301.01025},
 primaryClass = {astro-ph.HE},
       adsurl = {https://ui.adsabs.harvard.edu/abs/2023ApJ...944..103Z}
}

@ARTICLE{2025MNRAS.537.2380Z,
       author = {{Zhang}, Haiyun and {Yan}, Dahai and {Zhang}, Li and {Tang}, Niansheng},
        title = "{Evidence for magneto-gravitational processes in supermassive black hole binary PG 1553+113}",
      journal = {\mnras},
         year = 2025,
        month = mar,
       volume = {537},
       number = {3},
        pages = {2380-2386},
          doi = {10.1093/mnras/staf129},
archivePrefix = {arXiv},
       eprint = {2501.15169},
 primaryClass = {astro-ph.HE},
       adsurl = {https://ui.adsabs.harvard.edu/abs/2025MNRAS.537.2380Z}
}

@ARTICLE{2025ApJ...988..206Z,
       author = {{Zhang}, Haiyun and {Yan}, Dahai and {Zhang}, Li and {Tang}, Niansheng},
        title = "{Gaussian Process Modeling Coronal X-Ray Variability of Active Galactic Nuclei}",
      journal = {\apj},
         year = 2025,
        month = aug,
       volume = {988},
       number = {2},
          eid = {206},
        pages = {206},
          doi = {10.3847/1538-4357/adec92},
archivePrefix = {arXiv},
       eprint = {2507.04715},
 primaryClass = {astro-ph.HE},
       adsurl = {https://ui.adsabs.harvard.edu/abs/2025ApJ...988..206Z}
}

@ARTICLE{2021ApJ...919...58Z,
       author = {{Zhang}, Haiyun and {Yan}, Dahai and {Zhang}, Pengfei and {Yang}, Shenbang and {Zhang}, Li},
        title = "{A Quasi-periodic Oscillation in the {\ensuremath{\gamma}}-Ray Emission from the Non-blazar Active Galactic Nucleus PKS 0521-36}",
      journal = {\apj},
         year = 2021,
        month = sep,
       volume = {919},
       number = {1},
          eid = {58},
        pages = {58},
          doi = {10.3847/1538-4357/ac0cf0},
archivePrefix = {arXiv},
       eprint = {2106.10040},
 primaryClass = {astro-ph.HE},
       adsurl = {https://ui.adsabs.harvard.edu/abs/2021ApJ...919...58Z}
}

@ARTICLE{2025MNRAS.540.3790Z,
       author = {{Zhang}, Haiyun and {Yan}, Dahai and {Zhou}, Jianeng and {Zhang}, Li and {Tang}, Niansheng},
        title = "{Pattern and origin for the extreme {\ensuremath{\gamma}}-ray flares of 3C 454.3 and 3C 279: an astrophysical critical damper?}",
      journal = {\mnras},
         year = 2025,
        month = jul,
       volume = {540},
       number = {4},
        pages = {3790-3797},
          doi = {10.1093/mnras/staf944},
archivePrefix = {arXiv},
       eprint = {2502.20867},
 primaryClass = {astro-ph.HE},
       adsurl = {https://ui.adsabs.harvard.edu/abs/2025MNRAS.540.3790Z}
}

@ARTICLE{1995PhRvL..74.1060R,
       author = {{Rybicki}, George B. and {Press}, William H.},
        title = "{Class of Fast Methods for Processing Irregularly Sampled or Otherwise InhomogeneousOne-Dimensional Data}",
      journal = {\prl},
         year = 1995,
        month = feb,
       volume = {74},
       number = {7},
        pages = {1060-1063},
          doi = {10.1103/PhysRevLett.74.1060},
archivePrefix = {arXiv},
       eprint = {comp-gas/9405004},
 primaryClass = {nlin.CG},
       adsurl = {https://ui.adsabs.harvard.edu/abs/1995PhRvL..74.1060R}
}

@ARTICLE{1930PhRv...36..823U,
       author = {{Uhlenbeck}, G.~E. and {Ornstein}, L.~S.},
        title = "{On the Theory of the Brownian Motion}",
      journal = {Physical Review},
         year = 1930,
        month = sep,
       volume = {36},
       number = {5},
        pages = {823-841},
          doi = {10.1103/PhysRev.36.823},
       adsurl = {https://ui.adsabs.harvard.edu/abs/1930PhRv...36..823U}
}

@ARTICLE{1992ApJ...398..169R,
       author = {{Rybicki}, George B. and {Press}, William H.},
        title = "{Interpolation, Realization, and Reconstruction of Noisy, Irregularly Sampled Data}",
      journal = {\apj},
         year = 1992,
        month = oct,
       volume = {398},
        pages = {169},
          doi = {10.1086/171845},
       adsurl = {https://ui.adsabs.harvard.edu/abs/1992ApJ...398..169R}
}

@ARTICLE{2025arXiv250512671G,
       author = {{Guo}, Minghao and {Quataert}, Eliot and {Squire}, Jonathan and {Hopkins}, Philip F. and {Stone}, James M.},
        title = "{Idealized Global Models of Accretion Disks with Strong Toroidal Magnetic Fields}",
      journal = {arXiv e-prints},
         year = 2025,
        month = may,
          eid = {arXiv:2505.12671},
        pages = {arXiv:2505.12671},
          doi = {10.48550/arXiv.2505.12671},
archivePrefix = {arXiv},
       eprint = {2505.12671},
 primaryClass = {astro-ph.HE},
       adsurl = {https://ui.adsabs.harvard.edu/abs/2025arXiv250512671G}
}

@ARTICLE{2025MNRAS.543.1748M,
       author = {{Ma}, Ruican and {Done}, Chris and {Kubota}, Aya},
        title = "{Testing the Lense{\textendash}Thirring precession origin of the QPO in Swift J1727.8‑1613}",
      journal = {\mnras},
         year = 2025,
        month = oct,
       volume = {543},
       number = {2},
        pages = {1748-1760},
          doi = {10.1093/mnras/staf1524},
archivePrefix = {arXiv},
       eprint = {2506.18857},
 primaryClass = {astro-ph.HE},
       adsurl = {https://ui.adsabs.harvard.edu/abs/2025MNRAS.543.1748M}
}

@ARTICLE{2002A&A...387..487R,
       author = {{Rodriguez}, J. and {Varni{\`e}re}, P. and {Tagger}, M. and {Durouchoux}, Ph.},
        title = "{Accretion-ejection instability and QPO in black hole binaries I. Observations}",
      journal = {\aap},
         year = 2002,
        month = may,
       volume = {387},
        pages = {487-496},
          doi = {10.1051/0004-6361:20000524},
archivePrefix = {arXiv},
       eprint = {astro-ph/0203324},
 primaryClass = {astro-ph},
       adsurl = {https://ui.adsabs.harvard.edu/abs/2002A&A...387..487R}
}

@ARTICLE{2009MNRAS.397L.101I,
       author = {{Ingram}, Adam and {Done}, Chris and {Fragile}, P. Chris},
        title = "{Low-frequency quasi-periodic oscillations spectra and Lense-Thirring precession}",
      journal = {\mnras},
         year = 2009,
        month = jul,
       volume = {397},
       number = {1},
        pages = {L101-L105},
          doi = {10.1111/j.1745-3933.2009.00693.x},
archivePrefix = {arXiv},
       eprint = {0901.1238},
 primaryClass = {astro-ph.SR},
       adsurl = {https://ui.adsabs.harvard.edu/abs/2009MNRAS.397L.101I}
}

@ARTICLE{1973A&A....24..337S,
       author = {{Shakura}, N.~I. and {Sunyaev}, R.~A.},
        title = "{Black holes in binary systems. Observational appearance.}",
      journal = {\aap},
         year = 1973,
        month = jan,
       volume = {24},
        pages = {337-355},
       adsurl = {https://ui.adsabs.harvard.edu/abs/1973A&A....24..337S}
}

@ARTICLE{1996ApJ...463..656S,
       author = {{Stone}, James M. and {Hawley}, John F. and {Gammie}, Charles F. and {Balbus}, Steven A.},
        title = "{Three-dimensional Magnetohydrodynamical Simulations of Vertically Stratified Accretion Disks}",
      journal = {\apj},
         year = 1996,
        month = jun,
       volume = {463},
        pages = {656},
          doi = {10.1086/177280},
       adsurl = {https://ui.adsabs.harvard.edu/abs/1996ApJ...463..656S}
}

@ARTICLE{2010ApJ...713...52D,
       author = {{Davis}, Shane W. and {Stone}, James M. and {Pessah}, Martin E.},
        title = "{Sustained Magnetorotational Turbulence in Local Simulations of Stratified Disks with Zero Net Magnetic Flux}",
      journal = {\apj},
         year = 2010,
        month = apr,
       volume = {713},
       number = {1},
        pages = {52-65},
          doi = {10.1088/0004-637X/713/1/52},
archivePrefix = {arXiv},
       eprint = {0909.1570},
 primaryClass = {astro-ph.HE},
       adsurl = {https://ui.adsabs.harvard.edu/abs/2010ApJ...713...52D}
}

@ARTICLE{2011ApJ...730...94S,
       author = {{Simon}, Jacob B. and {Hawley}, John F. and {Beckwith}, Kris},
        title = "{Resistivity-driven State Changes in Vertically Stratified Accretion Disks}",
      journal = {\apj},
         year = 2011,
        month = apr,
       volume = {730},
       number = {2},
          eid = {94},
        pages = {94},
          doi = {10.1088/0004-637X/730/2/94},
archivePrefix = {arXiv},
       eprint = {1010.0005},
 primaryClass = {astro-ph.HE},
       adsurl = {https://ui.adsabs.harvard.edu/abs/2011ApJ...730...94S}
}

@ARTICLE{2013ApJ...767...30B,
       author = {{Bai}, Xue-Ning and {Stone}, James M.},
        title = "{Local Study of Accretion Disks with a Strong Vertical Magnetic Field: Magnetorotational Instability and Disk Outflow}",
      journal = {\apj},
         year = 2013,
        month = apr,
       volume = {767},
       number = {1},
          eid = {30},
        pages = {30},
          doi = {10.1088/0004-637X/767/1/30},
archivePrefix = {arXiv},
       eprint = {1210.6661},
 primaryClass = {astro-ph.HE},
       adsurl = {https://ui.adsabs.harvard.edu/abs/2013ApJ...767...30B}
}

@ARTICLE{2025OJAp....8E..39S,
       author = {{Squire}, Jonathan and {Quataert}, Eliot and {Hopkins}, Philip F.},
        title = "{Rapid, strongly magnetized accretion in the zero-net-vertical-flux shearing box}",
      journal = {The Open Journal of Astrophysics},
         year = 2025,
        month = apr,
       volume = {8},
          eid = {39},
        pages = {39},
          doi = {10.33232/001c.136467},
archivePrefix = {arXiv},
       eprint = {2409.05467},
 primaryClass = {astro-ph.HE},
       adsurl = {https://ui.adsabs.harvard.edu/abs/2025OJAp....8E..39S}
}

@ARTICLE{2016MNRAS.460.3488S,
       author = {{Salvesen}, Greg and {Armitage}, Philip J. and {Simon}, Jacob B. and {Begelman}, Mitchell C.},
        title = "{Strongly magnetized accretion discs require poloidal flux}",
      journal = {\mnras},
         year = 2016,
        month = aug,
       volume = {460},
       number = {4},
        pages = {3488-3493},
          doi = {10.1093/mnras/stw1231},
archivePrefix = {arXiv},
       eprint = {1602.04810},
 primaryClass = {astro-ph.HE},
       adsurl = {https://ui.adsabs.harvard.edu/abs/2016MNRAS.460.3488S}
}

@ARTICLE{1995ApJ...440..742H,
       author = {{Hawley}, John F. and {Gammie}, Charles F. and {Balbus}, Steven A.},
        title = "{Local Three-dimensional Magnetohydrodynamic Simulations of Accretion Disks}",
      journal = {\apj},
         year = 1995,
        month = feb,
       volume = {440},
        pages = {742},
          doi = {10.1086/175311},
       adsurl = {https://ui.adsabs.harvard.edu/abs/1995ApJ...440..742H}
}

@ARTICLE{2016MNRAS.457..857S,
       author = {{Salvesen}, Greg and {Simon}, Jacob B. and {Armitage}, Philip J. and {Begelman}, Mitchell C.},
        title = "{Accretion disc dynamo activity in local simulations spanning weak-to-strong net vertical magnetic flux regimes}",
      journal = {\mnras},
         year = 2016,
        month = mar,
       volume = {457},
       number = {1},
        pages = {857-874},
          doi = {10.1093/mnras/stw029},
archivePrefix = {arXiv},
       eprint = {1511.06368},
 primaryClass = {astro-ph.HE},
       adsurl = {https://ui.adsabs.harvard.edu/abs/2016MNRAS.457..857S}
}

@ARTICLE{2013MNRAS.428.2255P,
       author = {{Penna}, Robert F. and {S{\k{a}}dowski}, Aleksander and {Kulkarni}, Akshay K. and {Narayan}, Ramesh},
        title = "{The Shakura-Sunyaev viscosity prescription with variable {\ensuremath{\alpha}} (r)}",
      journal = {\mnras},
         year = 2013,
        month = jan,
       volume = {428},
       number = {3},
        pages = {2255-2274},
          doi = {10.1093/mnras/sts185},
archivePrefix = {arXiv},
       eprint = {1211.0526},
 primaryClass = {astro-ph.HE},
       adsurl = {https://ui.adsabs.harvard.edu/abs/2013MNRAS.428.2255P}
}

@ARTICLE{2024ApJ...961L..42Z,
       author = {{Zhao}, Qing-Chang and {Tao}, Lian and {Li}, Han-Cheng and {Zhang}, Shuang-Nan and {Feng}, Hua and {Ge}, Ming-Yu and {Ji}, Long and {Wang}, Ya-Nan and {Huang}, Yue and {Ma}, Xiang and {Zhang}, Liang and {Qu}, Jin-Lu and {Xu}, Yan-Jun and {Zhang}, Shu and {Yin}, Qian-Qing and {Shui}, Qing-Cang and {Ma}, Rui-Can and {Zhao}, Shu-Jie and {Li}, Pan-Ping and {Yang}, Zi-Xu and {Liu}, He-Xin and {Yu}, Wei},
        title = "{The First Polarimetric View on Quasiperiodic Oscillations in a Black Hole X-Ray Binary}",
      journal = {\apjl},
         year = 2024,
        month = feb,
       volume = {961},
       number = {2},
          eid = {L42},
        pages = {L42},
          doi = {10.3847/2041-8213/ad1e6c},
archivePrefix = {arXiv},
       eprint = {2401.08970},
 primaryClass = {astro-ph.HE},
       adsurl = {https://ui.adsabs.harvard.edu/abs/2024ApJ...961L..42Z}
}

@ARTICLE{2025arXiv251010353J,
       author = {{Jin}, Pei and {M{\'e}ndez}, Mariano and {Garc{\'\i}a}, Federico and {Altamirano}, Diego and {Vincentelli}, Federico M.},
        title = "{Black-hole X-ray binary Swift J1727.8$-$1613 shows simultaneous Type-B and Type-C quasi-periodic oscillations across the hard-intermediate and soft-intermediate states}",
      journal = {arXiv e-prints},
         year = 2025,
        month = oct,
          eid = {arXiv:2510.10353},
        pages = {arXiv:2510.10353},
          doi = {10.48550/arXiv.2510.10353},
archivePrefix = {arXiv},
       eprint = {2510.10353},
 primaryClass = {astro-ph.HE},
       adsurl = {https://ui.adsabs.harvard.edu/abs/2025arXiv251010353J}
}

@ARTICLE{2009MNRAS.396.1370F,
       author = {{Fender}, R.~P. and {Homan}, J. and {Belloni}, T.~M.},
        title = "{Jets from black hole X-ray binaries: testing, refining and extending empirical models for the coupling to X-rays}",
      journal = {\mnras},
         year = 2009,
        month = jul,
       volume = {396},
       number = {3},
        pages = {1370-1382},
          doi = {10.1111/j.1365-2966.2009.14841.x},
archivePrefix = {arXiv},
       eprint = {0903.5166},
 primaryClass = {astro-ph.HE},
       adsurl = {https://ui.adsabs.harvard.edu/abs/2009MNRAS.396.1370F}
}

@ARTICLE{2012MNRAS.421..468M,
       author = {{Miller-Jones}, J.~C.~A. and {Sivakoff}, G.~R. and {Altamirano}, D. and {Coriat}, M. and {Corbel}, S. and {Dhawan}, V. and {Krimm}, H.~A. and {Remillard}, R.~A. and {Rupen}, M.~P. and {Russell}, D.~M. and {Fender}, R.~P. and {Heinz}, S. and {K{\"o}rding}, E.~G. and {Maitra}, D. and {Markoff}, S. and {Migliari}, S. and {Sarazin}, C.~L. and {Tudose}, V.},
        title = "{Disc-jet coupling in the 2009 outburst of the black hole candidate H1743-322}",
      journal = {\mnras},
         year = 2012,
        month = mar,
       volume = {421},
       number = {1},
        pages = {468-485},
          doi = {10.1111/j.1365-2966.2011.20326.x},
archivePrefix = {arXiv},
       eprint = {1201.1678},
 primaryClass = {astro-ph.HE},
       adsurl = {https://ui.adsabs.harvard.edu/abs/2012MNRAS.421..468M}
}

@ARTICLE{1997ApJ...489..865E,
       author = {{Esin}, Ann A. and {McClintock}, Jeffrey E. and {Narayan}, Ramesh},
        title = "{Advection-Dominated Accretion and the Spectral States of Black Hole X-Ray Binaries: Application to Nova Muscae 1991}",
      journal = {\apj},
         year = 1997,
        month = nov,
       volume = {489},
       number = {2},
        pages = {865-889},
          doi = {10.1086/304829},
archivePrefix = {arXiv},
       eprint = {astro-ph/9705237},
 primaryClass = {astro-ph},
       adsurl = {https://ui.adsabs.harvard.edu/abs/1997ApJ...489..865E}
}

@ARTICLE{2023ATel16235....1K,
       author = {{Katoch}, Tilak and {Antia}, H.~M. and {Nandi}, Anuj and {Shah}, Parag},
        title = "{Detection of QPO in the peak of outburst of black hole candidate Swift J1727.8-1613: AstroSat/LAXPC Observation in Slew Mode}",
      journal = {The Astronomer's Telegram},
         year = 2023,
        month = sep,
       volume = {16235},
        pages = {1},
       adsurl = {https://ui.adsabs.harvard.edu/abs/2023ATel16235....1K}
}

@ARTICLE{2023ATel16243....1K,
       author = {{Katoch}, Tilak and {Nandi}, Anuj and {Shah}, Parag},
        title = "{Discovery detection of strong QPO at hard X-rays with AstroSat during 'unusual' peak of black hole transient Swift J1727.8-1613}",
      journal = {The Astronomer's Telegram},
         year = 2023,
        month = sep,
       volume = {16243},
        pages = {1},
       adsurl = {https://ui.adsabs.harvard.edu/abs/2023ATel16243....1K}
}

@ARTICLE{2024MNRAS.531.4893M,
       author = {{Mereminskiy}, I. and {Lutovinov}, A. and {Molkov}, S. and {Krivonos}, R. and {Semena}, A. and {Sazonov}, S. and {Tkachenko}, A. and {Sunyaev}, R.},
        title = "{Hard X-rays and QPO in Swift J1727.8-1613: the rise and plateau of the 2023 outburst}",
      journal = {\mnras},
         year = 2024,
        month = jul,
       volume = {531},
       number = {4},
        pages = {4893-4899},
          doi = {10.1093/mnras/stae1393},
archivePrefix = {arXiv},
       eprint = {2310.06697},
 primaryClass = {astro-ph.HE},
       adsurl = {https://ui.adsabs.harvard.edu/abs/2024MNRAS.531.4893M}
}

@ARTICLE{2025SCPMA..6819502Z,
       author = {{Zhang}, Shuang-Nan and {Santangelo}, Andrea and {Xu}, Yupeng and {Feng}, Hua and {Lu}, Fangjun and {Chen}, Yong and {Ge}, Mingyu and {Nandra}, Kirpal and {Wu}, Xin and {Feroci}, Marco and {Hernanz}, Margarita and {Liu}, Congzhan and {He}, Huilin and {Wang}, Yusa and {Jiang}, Weichun and {Cui}, Weiwei and {Yang}, Yanji and {Wang}, Juan and {Li}, Wei and {Li}, Hong and {Du}, Yuanyuan and {Liu}, Xiaohua and {Meng}, Bin and {Wen}, Xiangyang and {Zhang}, Aimei and {Ma}, Jia and {Li}, Maoshun and {Li}, Gang and {Qi}, Liqiang and {Sun}, Jianchao and {Luo}, Tao and {Liu}, Hongwei and {Liu}, Xiaojing and {Zhang}, Fan and {Luo}, Laidan and {Zhu}, Yuxuan and {Zhao}, Zijian and {Sun}, Liang and {Yang}, Xiongtao and {Wu}, Qiong and {Jiang}, Jiechen and {Shi}, Haoli and {Liu}, Jiangtao and {Xu}, Yanbing and {Yang}, Sheng and {Zhang}, Laiyu and {Han}, Dawei and {Gao}, Na and {Huo}, Jia and {Zhang}, Ziliang and {Wang}, Hao and {Zhao}, Xiaofan and {Wang}, Shuo and {Li}, Zhenjie and {Bao}, Ziyu and {Liu}, Yaoguang and {Wang}, Ke and {Wang}, Na and {Wang}, Bo and {Wang}, Langping and {Wang}, Dianlong and {Ding}, Fei and {Sheng}, Lizhi and {Qiang}, Pengfei and {Yan}, Yongqing and {Liu}, Yongan and {Wu}, Zhenyu and {Liu}, Yichen and {Chen}, Hao and {Zhang}, Yacong and {Liu}, Hongbang and {Altmann}, Alexander and {Bechteler}, Thomas and {Burwitz}, Vadim and {Fiorini}, Carlo and {Friedrich}, Peter and {Meidinger}, Norbert and {Strecker}, Rafael and {Baldini}, Luca and {Bellazzini}, Ronaldo and {Bonino}, Raffaella and {Frass{\`a}}, Andrea and {Latronico}, Luca and {Maldera}, Simone and {Manfreda}, Alberto and {Minuti}, Massimo and {Pesce-Rollins}, Melissa and {Sgr{\`o}}, Carmelo and {Tugliani}, Stefano and {Pareschi}, Giovanni and {Basso}, Stefano and {Sironi}, Giorgia and {Spiga}, Daniele and {Tagliaferri}, Gianpiero and {Tykhonov}, Andrii and {Paltani}, St{\`e}phane and {Bozzo}, Enrico and {Tenzer}, Christoph and {Bayer}, J{\"o}rg and {Tuo}, Youli and {Liu}, Honghui and {Zhang}, Yonghe and {Cai}, Zhiming and {Liu}, Huaqiu and {Chen}, Wen and {Wang}, Chunhong and {He}, Tao and {Chen}, Yehai and {Qiu}, Chengbo and {Zhang}, Ye and {Feng}, Jianchao and {Zhu}, Xiaofei and {Zhou}, Heng and {Zheng}, Shijie and {Song}, Liming and {Wang}, Jinzhou and {Jia}, Shumei and {Jiang}, Zewen and {Li}, Xiaobo and {Zhao}, Haisheng and {Guan}, Ju and {Zhang}, Juan and {Li}, Chengkui and {Huang}, Yue and {Liao}, Jinyuan and {You}, Yuan and {Zhang}, Hongmei and {Wang}, Wenshuai and {Wang}, Shuang and {Ou}, Ge and {Hu}, Hao and {Shi}, Jingyan and {Cui}, Tao and {Jiang}, Xiaowei and {Cheng}, Yaodong and {Li}, Haibo and {Xu}, Yanjun and {Zane}, Silvia and {Bambi}, Cosimo and {Bu}, Qingcui and {Dall'Osso}, Simone and {Rosa}, Alessandra De and {Gou}, Lijun and {Guillot}, Sebastien and {Ji}, Long and {Li}, Ang and {Mao}, Jirong and {Patruno}, Alessandro and {Stratta}, Giulia and {Taverna}, Roberto and {Tsygankov}, Sergey and {Uttley}, Phil and {Watts}, Anna L. and {Wu}, Xuefeng and {Xu}, Renxin and {Yi}, Shuxu and {Zhang}, Guobao and {Zhang}, Liang and {Zhao}, Wen and {Zhou}, Ping},
        title = "{The enhanced X-ray Timing and Polarimetry mission{\textemdash}eXTP for launch in 2030}",
      journal = {Science China Physics, Mechanics, and Astronomy},
         year = 2025,
        month = sep,
       volume = {68},
       number = {11},
          eid = {119502},
        pages = {119502},
          doi = {10.1007/s11433-025-2786-6},
archivePrefix = {arXiv},
       eprint = {2506.08101},
 primaryClass = {astro-ph.HE},
       adsurl = {https://ui.adsabs.harvard.edu/abs/2025SCPMA..6819502Z}
}

@ARTICLE{2025SCPMA..6819506Y,
       author = {{Yi}, Shu-Xu and {Zhao}, Wen and {Xu}, Ren-Xin and {Wu}, Xue-Feng and {Stratta}, Giulia and {Dall'Osso}, Simone and {Xu}, Yan-Jun and {Santangelo}, Andrea and {Zane}, Silvia and {Zhang}, Shuang-Nan and {Feng}, Hua and {Yang}, Huan and {Mao}, Junjie and {Ge}, Junqiang and {Shao}, Lijing and {Lan}, Mi-Xiang and {Gao}, He and {Lin}, Lin and {Jiang}, Ning and {Wu}, Qingwen and {Liu}, Tong and {Yu}, Yun-Wei and {Wang}, Xiang-Yu and {Zhang}, Jin and {Guetta}, Dafne and {Geng}, Jin-Jun and {Xiao}, Di and {Huang}, Yong-Feng and {Kang}, Yacheng and {Cao}, Tian-Yong and {Zhang}, Zhen and {Lyu}, Zhenwei and {Pan}, Zhen and {Chen}, Yunfeng and {Gao}, Yong and {Li}, Ang and {Fu}, Yu-Cong and {Xiao}, Shuo and {Wang}, Wei-Yang and {Wang}, Fayin and {Zhao}, Zhenyin and {Lei}, Weihua and {Shen}, Rong-Feng and {Dai}, Lixin and {Wu}, Guang-Lei and {Liu}, Liang-Duan and {Li}, Bing and {Fan}, Xilong and {Zhu}, Xing-Jiang and {Lu}, Youjun and {Xu}, Fan and {Cheng}, Kangfa and {Lin}, Da-Bin and {Zhao}, Xiao-Hong and {Wei}, Jun-Jie and {Zhang}, Bin-Bin and {Mao}, Ji-Rong and {Xue}, Yongquan and {Shu}, Xinwen and {Zhang}, Wenjie and {Lin}, Wei-Li and {Fiore}, Achille and {Li}, Zhuo and {Martin-Carrillo}, Antonio and {Fisher}, Joseph and {Xie}, Fei and {Li}, Ye and {Mereghetti}, Sandro and {Xiong}, Shao-Lin and {Yang}, Yu-Han and {Troja}, Eleonora and {Dai}, Zi-Gao and {Wei}, Da-Ming and {Liang}, En-Wei and {Horvath}, J.~E. and {Cunha Sampaio}, G.~R. and {Bar{\~a}o}, L.~G. and {de S{\'a}}, L.~M.},
        title = "{Prospects for time-domain and multi-messenger science with eXTP}",
      journal = {Science China Physics, Mechanics, and Astronomy},
         year = 2025,
        month = sep,
       volume = {68},
       number = {11},
          eid = {119506},
        pages = {119506},
          doi = {10.1007/s11433-025-2782-2},
archivePrefix = {arXiv},
       eprint = {2506.08368},
 primaryClass = {astro-ph.HE},
       adsurl = {https://ui.adsabs.harvard.edu/abs/2025SCPMA..6819506Y}
}

@ARTICLE{1997MNRAS.292..679L,
       author = {{Lyubarskii}, Yu. E.},
        title = "{Flicker noise in accretion discs}",
      journal = {\mnras},
         year = 1997,
        month = dec,
       volume = {292},
       number = {3},
        pages = {679-685},
          doi = {10.1093/mnras/292.3.679},
       adsurl = {https://ui.adsabs.harvard.edu/abs/1997MNRAS.292..679L}
}

@ARTICLE{2006MNRAS.367..801A,
       author = {{Ar{\'e}valo}, P. and {Uttley}, P.},
        title = "{Investigating a fluctuating-accretion model for the spectral-timing properties of accreting black hole systems}",
      journal = {\mnras},
         year = 2006,
        month = apr,
       volume = {367},
       number = {2},
        pages = {801-814},
          doi = {10.1111/j.1365-2966.2006.09989.x},
archivePrefix = {arXiv},
       eprint = {astro-ph/0512394},
 primaryClass = {astro-ph},
       adsurl = {https://ui.adsabs.harvard.edu/abs/2006MNRAS.367..801A}
}

@ARTICLE{2010MNRAS.404..738C,
       author = {{Cabanac}, C. and {Henri}, G. and {Petrucci}, P.-O. and {Malzac}, J. and {Ferreira}, J. and {Belloni}, T.~M.},
        title = "{Variability of X-ray binaries from an oscillating hot corona}",
      journal = {\mnras},
         year = 2010,
        month = may,
       volume = {404},
       number = {2},
        pages = {738-748},
          doi = {10.1111/j.1365-2966.2010.16340.x},
archivePrefix = {arXiv},
       eprint = {1001.2116},
 primaryClass = {astro-ph.HE},
       adsurl = {https://ui.adsabs.harvard.edu/abs/2010MNRAS.404..738C}
}

@ARTICLE{2003PASJ...55L..69N,
       author = {{Narayan}, Ramesh and {Igumenshchev}, Igor V. and {Abramowicz}, Marek A.},
        title = "{Magnetically Arrested Disk: an Energetically Efficient Accretion Flow}",
      journal = {\pasj},
         year = 2003,
        month = dec,
       volume = {55},
        pages = {L69-L72},
          doi = {10.1093/pasj/55.6.L69},
archivePrefix = {arXiv},
       eprint = {astro-ph/0305029},
 primaryClass = {astro-ph},
       adsurl = {https://ui.adsabs.harvard.edu/abs/2003PASJ...55L..69N}
}

@ARTICLE{2012ApJ...757L..20N,
       author = {{Narayan}, Ramesh and {{\"O}zel}, Feryal and {Sironi}, Lorenzo},
        title = "{Radio Synchrotron Emission from a Bow Shock around the Gas Cloud G2 Heading toward the Galactic Center}",
      journal = {\apjl},
         year = 2012,
        month = oct,
       volume = {757},
       number = {2},
          eid = {L20},
        pages = {L20},
          doi = {10.1088/2041-8205/757/2/L20},
archivePrefix = {arXiv},
       eprint = {1207.4215},
 primaryClass = {astro-ph.GA},
       adsurl = {https://ui.adsabs.harvard.edu/abs/2012ApJ...757L..20N}
}

@ARTICLE{2022ApJ...936...17H,
       author = {{H{\"u}bner}, Moritz and {Huppenkothen}, Daniela and {Lasky}, Paul D. and {Inglis}, Andrew R. and {Ick}, Christopher and {Hogg}, David W.},
        title = "{Searching for Quasi-periodic Oscillations in Astrophysical Transients Using Gaussian Processes}",
      journal = {\apj},
         year = 2022,
        month = sep,
       volume = {936},
       number = {1},
          eid = {17},
        pages = {17},
          doi = {10.3847/1538-4357/ac7959},
archivePrefix = {arXiv},
       eprint = {2205.12716},
 primaryClass = {astro-ph.IM},
       adsurl = {https://ui.adsabs.harvard.edu/abs/2022ApJ...936...17H}
}

@ARTICLE{2013MNRAS.431.3510S,
       author = {{Salvesen}, Greg and {Miller}, Jon M. and {Reis}, Rubens C. and {Begelman}, Mitchell C.},
        title = "{Spectral hardening as a viable alternative to disc truncation in black hole state transitions}",
      journal = {\mnras},
         year = 2013,
        month = jun,
       volume = {431},
       number = {4},
        pages = {3510-3532},
          doi = {10.1093/mnras/stt436},
archivePrefix = {arXiv},
       eprint = {1303.2116},
 primaryClass = {astro-ph.HE},
       adsurl = {https://ui.adsabs.harvard.edu/abs/2013MNRAS.431.3510S}
}

@ARTICLE{2015ApJ...813...84G,
       author = {{Garc{\'\i}a}, Javier A. and {Steiner}, James F. and {McClintock}, Jeffrey E. and {Remillard}, Ronald A. and {Grinberg}, Victoria and {Dauser}, Thomas},
        title = "{X-Ray Reflection Spectroscopy of the Black Hole GX 339--4: Exploring the Hard State with Unprecedented Sensitivity}",
      journal = {\apj},
         year = 2015,
        month = nov,
       volume = {813},
       number = {2},
          eid = {84},
        pages = {84},
          doi = {10.1088/0004-637X/813/2/84},
archivePrefix = {arXiv},
       eprint = {1505.03607},
 primaryClass = {astro-ph.HE},
       adsurl = {https://ui.adsabs.harvard.edu/abs/2015ApJ...813...84G}
}

@ARTICLE{2026ApJ..1003...73K,
       author = {{K{\"o}nig}, Ole and {Steiner}, James F. and {Bollemeijer}, Niek and {Connors}, Riley M.~T. and {Dauser}, Thomas and {Dov{\v{c}}iak}, Michal and {Fan}, Ningyue and {Garc{\'\i}a}, Javier A. and {Horn}, David and {Ingram}, Adam and {Lucchini}, Matteo and {Mastroserio}, Guglielmo and {Miller}, Cal and {Nathan}, Edward and {Nowak}, Michael A. and {Pottschmidt}, Katja and {Remillard}, Ron and {Song}, Yujia and {Svoboda}, Ji{\v{r}}{\'\i} and {van der Klis}, Michiel and {Ubach}, Santiago and {Wilms}, J{\"o}rn and {Zhang}, Yuexin},
        title = "{Systematic Assessment of Disk Truncation in the Black Hole X-Ray Binary Swift J1727.8─1613 Using NICER}",
      journal = {\apj},
         year = 2026,
        month = may,
       volume = {1003},
       number = {1},
          eid = {73},
        pages = {73},
          doi = {10.3847/1538-4357/ae60f5},
archivePrefix = {arXiv},
       eprint = {2604.14693},
 primaryClass = {astro-ph.HE},
       adsurl = {https://ui.adsabs.harvard.edu/abs/2026ApJ..1003...73K}
}

@ARTICLE{2026ApJ..1000..137C,
       author = {{Chand}, Swadesh and {Zdziarski}, Andrzej A. and {Dewangan}, Gulab C. and {Sahu}, Pragati},
        title = "{Evolution of the Inner Accretion Flow in Swift J1727.8─1613 across Intermediate States: Insights from Broadband Spectral and Timing Analysis}",
      journal = {\apj},
         year = 2026,
        month = mar,
       volume = {1000},
       number = {1},
          eid = {137},
        pages = {137},
          doi = {10.3847/1538-4357/ae45aa},
archivePrefix = {arXiv},
       eprint = {2512.05544},
 primaryClass = {astro-ph.HE},
       adsurl = {https://ui.adsabs.harvard.edu/abs/2026ApJ..1000..137C}
}

@ARTICLE{2024MNRAS.533.4188C,
       author = {{Carotenuto}, F. and {Fender}, R. and {Tetarenko}, A.~J. and {Corbel}, S. and {Zdziarski}, A.~A. and {Shaik}, G. and {Cooper}, A.~J. and {Di Palma}, I.},
        title = "{Constraining the physical properties of large-scale jets from black hole X-ray binaries and their impact on the local environment with blast-wave dynamical models}",
      journal = {\mnras},
         year = 2024,
        month = oct,
       volume = {533},
       number = {4},
        pages = {4188-4209},
          doi = {10.1093/mnras/stae2049},
archivePrefix = {arXiv},
       eprint = {2405.16624},
 primaryClass = {astro-ph.HE},
       adsurl = {https://ui.adsabs.harvard.edu/abs/2024MNRAS.533.4188C}
}

@ARTICLE{2026NewAR.10201746Z,
       author = {{Zdziarski}, Andrzej A. and {Marcel}, Gr{\'e}goire and {Veledina}, Alexandra and {Olejak}, Aleksandra and {Lan{\v{c}}ov{\'a}}, Debora},
        title = "{Spins of black holes in X-ray binaries and the tension with the gravitational wave measurements}",
      journal = {\nar},
         year = 2026,
        month = jun,
       volume = {102},
          eid = {101746},
        pages = {101746},
          doi = {10.1016/j.newar.2025.101746},
archivePrefix = {arXiv},
       eprint = {2506.00623},
 primaryClass = {astro-ph.HE},
       adsurl = {https://ui.adsabs.harvard.edu/abs/2026NewAR.10201746Z}
}

@ARTICLE{1945RvMP...17..323W,
       author = {{Wang}, Ming Chen and {Uhlenbeck}, G.~E.},
        title = "{On the Theory of the Brownian Motion II}",
      journal = {Reviews of Modern Physics},
         year = 1945,
        month = apr,
       volume = {17},
       number = {2-3},
        pages = {323-342},
          doi = {10.1103/RevModPhys.17.323},
       adsurl = {https://ui.adsabs.harvard.edu/abs/1945RvMP...17..323W}
}

\appendix

\section{Binned Lomb--Scargle power spectral density}
\label{appendix:binned_lsp}

\begin{figure}[!t]
    \centering
    \includegraphics[width=\columnwidth]{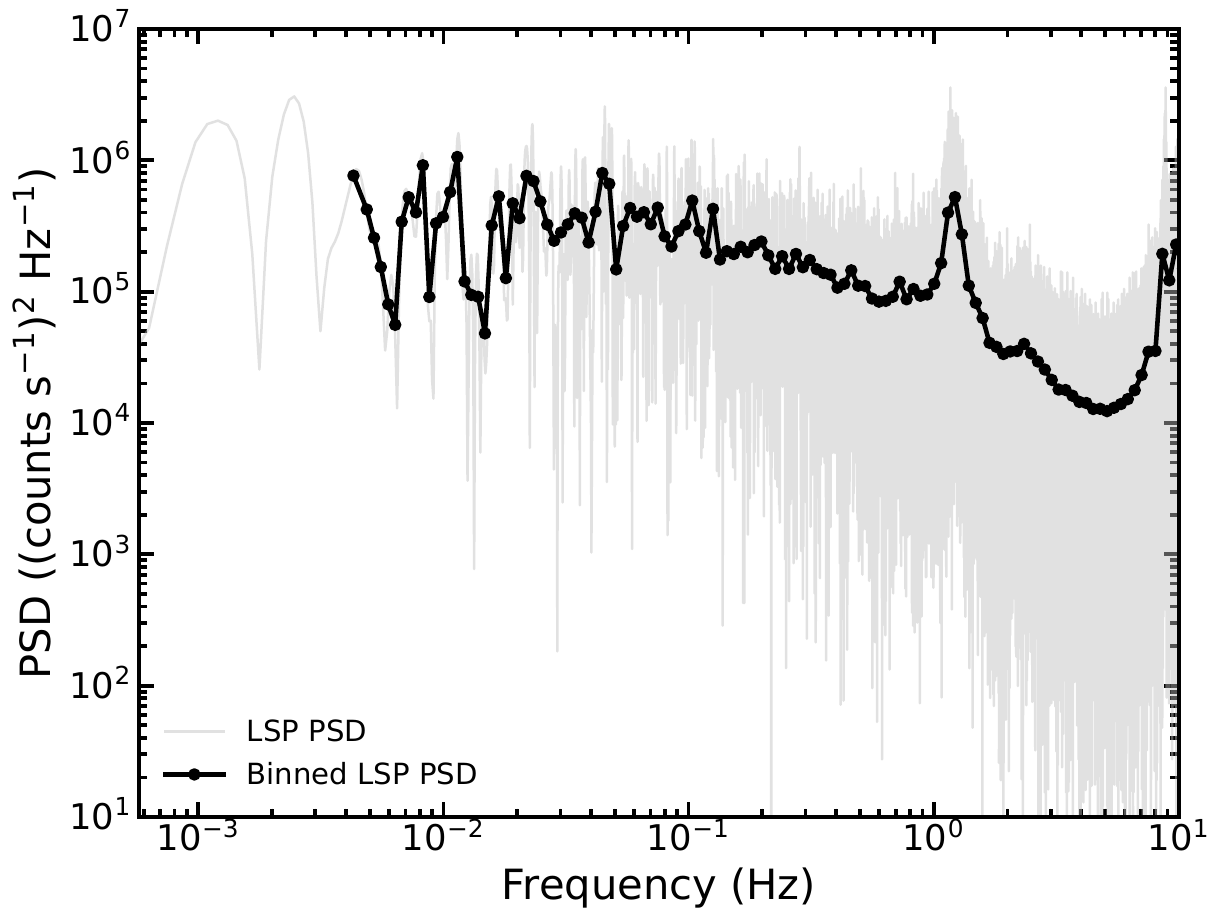}
    \caption{
    {
    Comparison between the LSP PSD of the light curve (gray line) and the logarithmically binned LSP PSD (black data points).
    }
    }
    \label{fig:binned_lsp}
\end{figure}

To improve the visualization of the power-spectral features shown in Fig.~\ref{Fig3}, we constructed a frequency-binned version of the Lomb--Scargle periodogram (LSP) power spectral density, as presented in Fig.~\ref{fig:binned_lsp}. We first calculated the LSP PSD using the original frequency grid adopted in the main analysis. The full frequency range was then divided into logarithmically spaced bins. For each bin, we calculated the arithmetic mean of the PSD values contained within that bin. The representative frequency was defined as the logarithmic center of the corresponding frequency interval. Bins containing too few frequency samples to provide a reliable average were excluded.

\end{document}